\documentclass[showpacs,aps,prd,nofootinbib,floatfix,amsmath,amssymb]{revtex4-1}
\usepackage{graphicx}
\usepackage{color}
\usepackage{enumerate} 
\usepackage{multirow}

\def\beq{\begin{eqnarray}}
\def\eeq{\end{eqnarray}}

\usepackage{slashed}

\begin{document}

\title{Charged Higgs boson production via $cb-$fusion at the Large Hadron Collider }
\author{J. Hern\'andez-S\'anchez}
\email{jaime.hernandez@correo.buap.mx}
\affiliation{Fac. de Cs. de la Electr\'onica, Benem\'erita Universidad Aut\'onoma de Puebla, Apartado Postal 1152, 72570 Puebla, Puebla, M\'exico}
\author{C. G. Honorato }
\email{carlosg.honorato@correo.buap.mx}
\affiliation{Fac. de Cs. de la Electr\'onica, Benem\'erita Universidad Aut\'onoma de Puebla, Apartado Postal 1152, 72570 Puebla, Puebla, M\'exico}

\author{S. Moretti}
\email{s.moretti@soton.ac.uk}
\affiliation{School of Physics and Astronomy, University of Southampton, Highfield, Southampton SO17 1BJ, United Kingdom, and Particle Physics Department, Rutherford Appleton Laboratory, Chilton, Didcot, Oxon OX11 0QX, United Kingdom}
\author{S. Rosado-Navarro}
\email{sebastian.rosado@gmail.com}
\affiliation{Fac. de Cs. F\'{\i}sico-Matem\'aticas, Benem\'erita Universidad Aut\'onoma de Puebla, Apartado Postal 1364, C.P.  72570 Puebla, Puebla, M\'exico}


\date{\today}

\begin{abstract}
We  analyse the production of a light charged Higgs boson at the  Large Hadron  Collider (LHC) via the quark-fusion mechanism $c \bar{b} \to H^- $  considering the decay channel  $H^- \to \tau \bar \nu_\tau$ in the final state. We study this process in the framework of the 2-Higgs Doublet Model Type III (2HDM-III) which assumes a four-zero texture in the Yukawa matrices and a general Higgs potential, wherein the two Higgs doublets coupling to both up and down fermions do generate Flavour Changing Neutral Currents (FCNCs) yet the latter can be controlled by the texture when flavour physics constraints are considered. We consider the parameter space of the model where this signal is enhanced and in agreement with both theoretical constraints and experimental data. In particular, we exploit  the setup with lepton-specific-like Yukawa couplings and  assess the LHC sensitivity to such $H^\pm$ signals against the dominant {irreducible and} reducible backgrounds. We show that   in our model   BR$(H^\pm \to cb) \sim 0.1- 0.2$ and BR$(H^\pm \to \tau \nu) \sim 0.7-0.9$  so that, under these  conditions,  the prospects for $H^\pm$ detection in the 2HDM-III in the aforementioned production and decay channels are excellent assuming standard collider energy and luminosity conditions.   
\end{abstract}

\maketitle

\section{Introduction}

In July 2012,  at the Large Hadron Collider (LHC),  a neutral spinless boson was discovered   by both  the ATLAS \cite{ATLAS} and CMS \cite{CMS} collaborations. This new state of Nature is very compatible  with the Standard Model (SM) Higgs boson, so this theoretical construct seems to be fully established now.  
However, the SM-like limit of Electro-Weak Symmetry Breaking (EWSB) dynamics induced by a Higgs potential exists in several Beyond the SM (BSM) extensions of the Higgs sector. Notably,  the 
2-Higgs Doublet Model (2HDM) \cite{Branco:2011iw}
in its  Types I, II, III (or Y)   and IV (or X), wherein Flavour Changing Neutral Currents (FCNCs) mediated by (pseudo)scalar Higgs states can be eliminated under discrete symmetries \cite{Branco:2011iw}, is an intriguing  BSM candidate, owing to the fact that it implements the same fundamental doublet structure of the SM (in fact,  twice), assumes the same SM gauge symmetry group  (i.e., $SU(3)_C\times SU(2)_L\times U(1)_Y$)  and predicts a variety of new Higgs boson signatures  that may be accessible at the LHC. In particular, of the eight degrees of freedom pertaining to a 2HDM, upon EWSB giving mass to the $W^\pm$ and $Z$ bosons, five  survive as physical Higgs bosons: three are neutral (two CP-even, $h$ and $H$ with, conventionally, $M_h<M_H$ plus one CP-odd, $A$) while two are charged ($H^\pm$).

However, another, equally  interesting kind of  2HDM is  the one where FCFNs can be controlled by a particular texture in the Yukawa matrices  \cite{Fritzsch:2002ga}. In particular, in previous papers, we have implemented a four-zero texture in a scenario which we have called 2HDM Type III (2HDM-III) \cite{DiazCruz:2009ek}. This model has a phenomenology that is very rich, which we studied at colliders in various instances \cite{Hernandez-Sanchez:2016vys}--\cite{HernandezSanchez:2013xj}, and some very interesting aspects, like flavour-violating quarks decays, which can be enhanced for  neutral Higgs bosons with intermediate mass (i.e., below  twice the $Z$ boson mass). 

Furthermore, in this model, the parameter space can avoid many of the current experimental constraints from flavour  and Higgs physics and   a light charged Higgs  boson  (i.e., with a mass below  the top quark one) is allowed therein \cite{HernandezSanchez:2012eg}, so that  the  decay $H^- \to b \bar{c}$ is enhanced and its Branching Ratio
 (BR) can be dominant, above and beyond those of the customary (flavour diagonal) $s\bar c$ and $\tau\nu$ channels. (In fact, this channel  has been also studied in a variety of Multi-Higgs Doublet Models (MHDMs) \cite{Akeroyd:2016ymd,Akeroyd:2012yg}, wherein the BR$ (H^- \to b \bar{c}) \approx 0.7- 0.8$ and one could obtain a considerable gain in sensitivity to the presence of a $H^-$  by tagging the $b$ quark.)  Finally, we have also performed a study of the  process $e^-p \to \nu_e H^- b $ followed by the signal $H^- \to b\bar{c} $ \cite{Hernandez-Sanchez:2016vys,Flores-Sanchez:2018dsr,Flores-Sanchez:2019jcx} at the Large Hadron electron Collider (LHeC), finding good detection prospects. 

In this work, by exploiting the enhancement of the $H^- \to c\bar{b}$ vertex and  building on the results previously presented in \cite{HernandezSanchez:2012eg}, we study the production of a light charged Higgs boson at the  LHC  via heavy-quark fusion, $b \bar{c} \to H^- $, followed by the decay  $H^- \to \tau \bar \nu_\tau$ (hereafter, c.c. channels are always  implied). We investigate these processes in the framework of the aforementioned 2HDM-III with so-called lepton-specific couplings  and assess the LHC sensitivity to this  production and decay dynamics  against the leading background, i.e., the irreducible one $q \bar{q}'  \to W^- \to \tau \bar \nu_\tau$, and the reducible noise produced by $g {q}'  \to W^{\pm} q$ (with an additional jet)
and $q \bar{q}  \to W^{+} W^{-} \to l^{+}l^{-}\nu \nu$ (where one lepton escapes detection, given that we will be looking for
leptonic decays of the $\tau$ in the signal. 
An up-to-date overview of charged Higgs boson phenomenology at the LHC can be found in Refs.~\cite{Akeroyd:2016ymd,Arhrib:2018ewj}. 

The plan of this paper is as follows. In the next section we describe the 2HDM-III. Then we introduce some benchmark configurations of it for the purpose of running a  Monte Carlo (MC) simulation and discussing the ensuing signal and background results. Finally, we conclude.

\section{The 2HDM-III}

In the 2HDM-III there are two (pseudo)scalar Higgs doublets, $\Phi_1^\dag=( \phi_{1}^{-},\phi_{1}^{0*} )$ and $\Phi_{2}^{\dag}=(\phi_{2}^{-},\phi_{2}^{0*})$, with hypercharge +1, and both couple to all fermions. In order to control FCNCs, as intimated, we have implemented a specific four-zero texture as an effective flavour theory in the Yukawa sector, so that a discrete symmetry is not necessary  \cite{Felix-Beltran:2013tra,HernandezSanchez:2012eg}. Then the  $SU(2)_L \times U(1)_Y$ invariant scalar potential should be the most general one:  
\begin{eqnarray}
V(\Phi_1,\Phi_2) &=& \mu_{1}^{2}(\Phi_{1}^{\dag}\Phi_{1}^{}) + \mu_{2}^{2}(\Phi_{2}^{\dag}\Phi_{2}^{}) - \left(\mu_{12}^{2}(\Phi_{1}^{\dag}\Phi_{2}^{} + h.c.)\right) \nonumber \\ 
&+& \frac{1}{2} \lambda_{1}(\Phi_{1}^{\dag}\Phi_{1}^{})^2 + \frac{1}{2} \lambda_{2}(\Phi_{2}^{\dag}\Phi_{2}^{})^2 + \lambda_{3}(\Phi_{1}^{\dag}\Phi_{1}^{})(\Phi_{2}^{\dag}\Phi_{2}^{}) + \lambda_{4}(\Phi_{1}^{\dag}\Phi_{2}^{})(\Phi_{2}^{\dag}\Phi_{1}^{}) \nonumber \\
&+& \left( \frac{1}{2} \lambda_{5}(\Phi_{1}^{\dag}\Phi_{2}^{})^2  + \lambda_{6}(\Phi_{1}^{\dag}\Phi_{1}^{})(\Phi_{1}^{\dag}\Phi_{2}^{}) + \lambda_{7}(\Phi_{2}^{\dag}\Phi_{2}^{})(\Phi_{1}^{\dag}\Phi_{2}^{}) + h.c. \right).
\end{eqnarray}
Here, we have assumed all parameters to be real, including the Vacuum Expectation Values (VEVs) of the (pseudo)scalar fields, therefore there is no CP-Violating (CPV) dynamics. Furthermore, note that, typically, the   $\lambda_6$ and $\lambda_7$ parameters are absent when a discrete symmetry is considered (e.g., $\Phi_1 \to \Phi_1$ and $\Phi_2 \to -\Phi_2$ ). 

Other than the physical Higgs masses ($M_h, M_H, M_A$ and $ M_H^\pm$), further independent  parameters of the 2HDM are the mixing angles $\alpha$ (related to the mass matrix of the CP-even sector) and $\beta$ (where $\tan \beta $ is the ratio of the two VEVs of the 2HDM). In our model, 2HDM-III, a four-zero texture is implemented as the mechanism that controls FCNCs and  the terms proportional to 
$\lambda_6$ and $\lambda_7$ are kept.  Herein, the EW parameter  $\rho = M_W^2/ M_Z^2 \cos_W^2$  can receive corrections at one-loop level proportional to the difference between the charged Higgs and CP-even/odd masses, but it is not sensitive to the value of $\lambda_6$ and $\lambda_7$ \cite{Cordero-Cid:2013sxa}. In particular,  when the difference of the scalars masses $M_{H^\pm} -M_{A}$($M_{H^\pm} -M_{H}$) is large, the subjacent custodial symmetry (twisted custodial symmetry)  is broken. Then, a survival model  to this EW observable is realised when $\rho \approx 1$   \cite{Gunion:2002zf,Gerard:2007kn,deVisscher:2009zb}.  In general, the above mass splitting appears also  in the expressions of the oblique parameters $S$, $T$ and $U$ (the so-called EW Precisions Observables (EWPOs)) \cite{Kanemura:2011sj}, so they should be reconciled too with the corresponding experimental bounds \cite{pdg:2018}. Hence, the benchmark scenarios chosen for our model  in the next section will be in agreement with these EW measurements.

For our model the Yukawa Lagrangian  is given by  \cite{HernandezSanchez:2012eg}:
\begin{equation}
\label{yuklan} 
\mathcal{L}_Y = -\left( Y_{1}^{u} \bar{Q}_{L} \tilde{\Phi}_{1} u_{R} + Y_{2}^{u} \bar{Q}_{L} \tilde{\Phi}_{2} u_{R} + Y_{1}^{d} \bar{Q}_{L} \Phi_{1} d_{R} + Y_{2}^{d} \bar{Q}_{L} \Phi_{2} d_{R} + Y_{1}^{l} \bar{L}_{L} \tilde{\Phi}_{1} l_{R} + Y_{2}^{l} \bar{L}_{L} \tilde{\Phi}_{2} l_{R} \right),
\end{equation}
where  $\tilde{\Phi}_{1,2} = i\sigma_2 \Phi_{1,2}^{*}$.  The fermion mass matrices after EWSB are: $M_f = \tfrac{1}{\sqrt{2}} \left( v_1 Y_{1}^{f} + v_2 Y_{2}^{f} \right),$ $f=u,d,l$, and both Yukawa matrices $Y_{1}^{f}$ and $Y_{2}^{f}$ have the aforementioned four-zero texture form and are Hermitian. 
Once  diagonalisation is done, $\bar{M}_f = V_{fL}^{\dag} M_f V_{fR}$, with $\bar{M}_f = \tfrac{1}{\sqrt{2}} \left( v_1 \tilde{Y}_{1}^{f} + v_2 \tilde{Y}_{2}^{f} \right),$ and $\tilde{Y}_{i}^{f} = V_{fL}^{\dag} Y_{i}^{f} V_{fR}$, we can get from the product $V_{q} Y_{n}^{q} V_{q}^{\dag}$  the rotated matrix $\tilde{Y}_{n}^{q}$ as  \cite{HernandezSanchez:2012eg}:
\begin{equation}
\left[ \tilde{Y}_{n}^{q} \right]_{ij} = \frac{\sqrt{m_{i}^{q} m_{j}^{q}}}{v} \left[ \tilde{\chi}_{n}^{q} \right]_{ij} = \frac{\sqrt{m_{i}^{q} m_{j}^{q}}}{v} \left[ \chi_{n}^{q} \right]_{ij} e^{i\vartheta_{ij}^{q}} ,
\end{equation}
where the $\chi$s are unknown dimensionless parameters of the model. Following  the procedure of \cite{HernandezSanchez:2012eg}, one can get the interactions of the charged Higgs bosons with the fermions,
\begin{eqnarray}
\mathcal{L}^{\bar{f_i}f_j\phi}= &-& \left\lbrace \frac{\sqrt{2}}{v} \bar{u}_i \left( m_{d_j} X_{ij} P_R + m_{u_i} Y_{ij} P_L \right) d_j H^{+} + \frac{\sqrt{2} m_{l_j}}{v} Z_{ij} \bar{\nu}_L l_R H^{+} + h.c. \right\rbrace, 
\end{eqnarray}
where 
$X_{ij}$, $Y_{ij}$ and $Z_{ij}$ are defined as follows \footnote{Hereafter, $V_{\rm CKM}$ is the Cabibbo-Kobayashi-Maskawa matrix.}:
\begin{eqnarray}
X_{ij} &=& \sum_{l=1}^{3} \left( V_{\rm CKM} \right)_{il} \left[ X \frac{m_{d_l}}{m_{d_j}} \delta_{lj} -\frac{f(X)}{\sqrt{2}} \sqrt{\frac{m_{d_l}}{m_{d_j}}} \tilde{\chi}_{lj}^{d} \right], \\
Y_{ij} &=& \sum_{l=1}^{3} \left[ Y \delta_{il} -\frac{f(Y)}{\sqrt{2}} \sqrt{\frac{m_{u_l}}{m_{u_i}}} \tilde{\chi}_{il}^{u} \right] \left( V_{\rm CKM} \right)_{lj}, \\
Z_{ij}^{l} &=& \left[ Z \frac{m_{l_i}}{m_{l_j}} \delta_{ij} -\frac{f(Z)}{\sqrt{2}} \sqrt{\frac{m_{l_i}}{m_{l_j}}} \tilde{\chi}_{ij}^{l} \right],
\end{eqnarray}
where 
$f(a)=\sqrt{1+a^2}$  and the parameters $X$, $Y$ and $Z$   are arbitrary complex numbers that can be linked to $\tan \beta$ or $\cot \beta$ when $\chi_{ij}^f=0$ \cite{HernandezSanchez:2012eg}, so that it  is then possible to  recover the standard four types  of  2HDM (see  Tab.  \ref{XYZ})\footnote{Hence, we will refer to these 2HDM-III `incarnations' as 2HDM-III like-$\chi$ scenarios, where $\chi=$ I, II, X and Y.}. Furthermore, the Higgs-fermion-fermion couplings $(\phi ff)$  in the 2HDM-III can be written as $g_{\rm 2HDM-III}^{\phi ff} = g_{\rm 2HDM-any}^{\phi ff} + \Delta g$, where $g_{\rm 2HDM-any}^{\phi ff}$ is the coupling $\phi f f$ in any of the 2HDMs with discrete symmetry and $\Delta g$ is the contribution of the four-zero texture. Lastly, we  also  point out that this Lagrangian can represent a Multi-Higgs Doublet Model (MHDM) or an Aligned 2HDM (A2HDM) with additional flavour physics in the Yukawa matrices \cite{Felix-Beltran:2013tra,HernandezSanchez:2012eg}. \begin{table}[htp]
\begin{center}
\begin{tabular}{|c|c|c|c|}
\hline
2HDM-III & $X$ & $Y$& $Z$\\
\hline
2HDM Type I & $-\cot \beta$ &  $ \cot \beta$ & $- \cot \beta$ \\
\hline
2HDM Type II & $\tan \beta$ &  $ \cot \beta$ & $ \tan \beta$ \\
\hline
2HDM Type X & $-\cot \beta$ &  $ \cot \beta$ & $ \tan \beta$ \\
\hline
2HDM Type Y  & $\tan \beta$ &  $ \cot \beta$ & $ -\cot \beta$ \\
\hline
\end{tabular}
\end{center}
\caption{The parameters $X$, $Y$ and $ Z$ of the 2HDM-III defined in the Yukawa interactions when $\chi_{ij}^f=0$ so as to recover the standard four types of 2HDM.}
\label{XYZ}
\end{table}%

\section{Benchmark scenario} \label{BP}

We have constrained our model using flavour and Higgs physics (i.e., the measurements of the SM-like Higgs boson discovered at the LHC plus the exclusions emerging from void searches for additional Higgs states at any collider)
 as well as EWPOs and theoretical bounds (like vacuum stability, unitarity and perturbativity).  
While we do not discuss the theoretical constraints (as they are a simple application of textbook methods), we dwell here at some length on all the experimental ones, with the intent of emphasising those applicable to a charged Higgs state.

Specifically,  the model is found in agreement with  flavour physics constraints by taking into account the analyses performed in  Refs. \cite{Felix-Beltran:2013tra,HernandezSanchez:2012eg,Crivellin:2013wna}, where the parameter space of the 2HDM-III is constrained by leptonic and semi-leptonic meson decays, like the  inclusive decays $B \to X_s \gamma$, $B_0-B_0$ as well as $K_0-K_0 $ mixing and $ B_s \to \mu^+ \mu^-$ transitions. Here, the Yukawa texture used in the model plays a relevant role in the amplitudes of the mesonic decays, altogether allowing for the possibility to obtain a light charged Higgs state, of order 100 GeV or so, in the case of Type X couplings (with all other Yukawa cases being more constrained in terms of $M_{H^\pm}$).  

Further, as for constraints from the SM-like Higgs boson measurements, 
we consider the impact at one loop-level of charged Higgs bosons on the radiative decays $h\to \gamma \gamma$ and $ \gamma Z$, as detailed in \cite{Cordero-Cid:2013sxa}. For this analysis, some of the most recent experimental data from the LHC are considered, namely, from Refs.  \cite{Sirunyan:2018tbk,Sirunyan:2018koj,Aaboud:2017uhw,Aaboud:2018ezd,Aaboud:2018wps}.  Once again, the Yukawa texture is involved in the couplings of the charged Higgs boson with fermions in the loop and low masses for a Type X Yukawa structure are allowed.

As for the current bounds on the mass of a charged Higgs boson from direct searches at present and past colliders, we have considered the following, recalling that for a light charged Higgs boson the main production mode at lepton machines is via $e^+e^-\to H^+H^-$  while  at hadron colliders is via $gg\to t\bar b H^-$  + c.c. (so that, for $M_{H^\pm}<m_t$, the latter correspond to top pair production and decay via a charged Higgs boson, i.e., $t\to bH^+$).  
 \begin{itemize}
 
\item LEP limits.  For the mass of the charged Higgs boson, the LEP collaborations have finally established  a universal lower bound at 78.6 GeV \cite{Schael:2006cr}.

 \item Tevatron limits. For a charged Higgs boson with a mass between 90 GeV and 160 GeV, CDF and D0 established a bound for the the BR$(t\to b  H^+ )$ of $ \approx 20 \%$  taking BR$(H^+\to c\bar{s})=1$ or  BR$(H^+\to \tau^+ \nu)=1$ \cite{Abbott:1999eca,Abulencia:2005jd,Abazov:2009aa}.

\item LHC limits. For the case BR$(H^+\to \tau^+ \nu)=1$ in the range of masses varying from  80 GeV to 160 GeV,  the CMS experiment has established  a BR$(t\to b H^+)=2-3\%$ as upper limit. Meanwhile,  for the mass range 90 GeV to 160 GeV with BR$(H^+\to c\bar{s})=1$, both ATLAS and CMS set BR$(t\to bH^+)\approx 20 \%$ as a maximum \cite{pdg:2018}. Finally, assuming BR$(H^+\to c\bar{b})=1$, in the mass range 90 GeV to 150 GeV, the CMS collaboration has set an upper limit of  BR$(t\to H^+ b)=0.5-0.8\%$ \cite{Sirunyan:2018dvm}.

 \end{itemize}

As for EW data, we have fixed the oblique parameter $U=0$, because this  is suppressed with respect  to the parameters $S$ and $T$ when a scale for new physics (just) above the EW regime is considered  \cite{pdg:2018}, taking $S= 0.02 \pm 0.07$ and $ T=0.06\pm 0.06$. 

Upon the application of all limits above, the following parameter space region roughly survives and is analysed here: 
$M_h =125$ GeV (thus with $h$ being the SM-like Higgs boson), $M_A = 100$ GeV, 180 GeV $< M_H < 260$ GeV and 100 GeV $<M_{H^\pm}<$ 170 GeV, with $0.1<\cos (\beta - \alpha)<0.5$.  Over such an expanse of parameter space, we  consider four scenarios, each in turn being an incarnation of our 2HDM-III:  like-I (where one Higgs doublet couples to all fermions); like-II (where one Higgs doublet couples to the up-type quarks and the other to the down-type quarks);  like-X (also called IV or "Lepton-specific", where the quark couplings are Type I and the lepton ones are Type II); like-Y (also called III or "Flipped", where the quark couplings are Type II and the lepton ones are Type I). 

For a light charged Higgs boson, in the 2HDM-III,  the most important decay channels are $H^\pm \to s c$ and $ b  c$, when $Y \gg X, Z$ (like-I scenario), $X, \, Z \gg Y$ (like-II scenario) or $X \gg Y, \, Z$  (like-Y scenario), in which cases the mode $H^\pm \to b c$ receives  a substantial enhancement coming from the  four-zero texture implemented in the Yukawa matrices, so one can even get  a BR$(H^- \to b\bar c) \approx 0.95$. However, this does not happen for $H^\pm \to \tau \nu$, which is the decay we must rely on in order to extract a charged Higgs boson signal in the hadronic environment  of the
LHC, specifically, assuming a leptonic decay of the $\tau$. For the case $Z \gg X, Y$ (like-X scenario),  the  decay channel $H^-  \to \tau \bar\nu_\tau$ is maximised, reaching a BR of 90\%  or so \cite{HernandezSanchez:2012eg}, 
while not penalising the  $H^\pm \to b c$ mode excessively, so that, in turn,  the production $c\bar{b}  \to H^+$ can reach a considerable cross section. In fact, a typical configuration is  BR$(H^\pm \to \tau \nu) \approx 0.9$ and BR$(H^\pm \to cb) \approx 0.1$. Guided by the parameter scan performed in \cite{Das:2015kea}, we finally adopt the following Benchmark Point (BP) in order to analyse by MC simulation at the LHC the  process  
$c\bar b  \to H^+  \to \tau \bar\nu_\tau$, as it offers the most optimistic chances for detection.
\begin{itemize}
\item Scenario 2HDM-III like-X: $\cos (\beta- \alpha)= 0.5$, $\chi^u_{22}= 1$, $\chi^u_{23}= 0.1$, $\chi^u_{33}= 1.4$, $\chi^d_{22}= 1.8$, $\chi^d_{23}= 0.1$,
$\chi^d_{33}= 1.2$,  $\chi^\ell_{22} =-0.4, \chi^\ell_{23}= 0.1$, $\chi^\ell_{33} =1$ with 
$Z \gg X, \, Y$.
Further, we assume $M_h=125$ GeV,  $M_A =100$ GeV,  $ M_H = 150$ GeV and 100 GeV $<M_{H^\pm}<$ 170 GeV. In fact, eventually, given the significant signal-to-background rates obtained for a light charged Higgs boson state, we will push our analysis up to 1 TeV or so for its mass.
\end{itemize}

\section{Numerical results}

As already stressed, we will attempt to establish the signal $b \bar{c} \to H^-  \to \tau \bar \nu_\tau$ at the LHC, by surpassing the results of Ref. \cite{HernandezSanchez:2012eg}, wherein a similar analysis was performed, although over a region of parameter space of the 2HDM-III which has largely been ruled out since, following the subsequent discovery of a SM-like Higgs boson at the LHC as well as the measurements of its properties therein. In fact, since that paper, also a myriad of void experimental searches for additional Higgs bosons were carried out by the LHC collaborations, which also impinge on the available 2HDM-III parameter space.  

\begin{figure}
\centering
\includegraphics[scale=0.4]{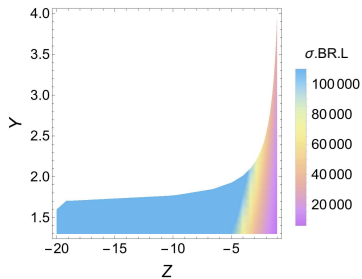}
\caption{Event rates for our BP at parton-level for $M_{H^\pm}= 120 $ GeV and $X=-1/Z$, assuming $\sqrt s=13$ TeV and $L=36.1$ fb$^{-1}$.}
\label{BP-X}
\end{figure}

As intimated, the cross section for our signal process is too small in the 2HDM-III incarnations of Type I,  II and  Y, therefore only 
the Type X realisation  is explored here. It was seen in our scan that its value is maximised for small $X$, so we fixed the latter to be $X=-1/Z$. The results of our scan over the plane $(Y,Z)$ are presented in Fig.~\ref{BP-X}, in terms of the $\sigma(b \bar{c} \to H^- )$ $\times~{\rm BR}(H^-  \to \tau \bar \nu_\tau)$ $\times~L$ yield\footnote{In fact, we use here a factorisation formula exploiting the charged Higgs boson in Narrow Width Approximation (NWA), given that it is very narrow. (This is done for calculation efficiency purposes.)}, where $L=36.1$ fb$^{-1}$ is the LHC luminosity at an energy of $\sqrt s=13$ TeV, corresponding to the values  used by the CMS collaboration in their $H^{\pm} \rightarrow \tau^{\pm}\nu_\tau$ decay channel analysis \cite{Sirunyan:2019hkq}. Here, we fix $M_{H^\pm}= 120 $ GeV for reference. It is clear that the inclusive rate  is very significant, the best point being $X=0.04$, $Y=1.6$, $Z=-20$, which produces  $\approx 2.276 \times 10^{6}$ events.

 In order to carry out our numerical analysis, we have used CalcHEP 3.7 \cite{Belyaev:2012qa}  as parton level event generator, interfaced to the CTEQ6L1 Parton Distribution Functions (PDFs) \cite{Pumplin:2002vw}  and  to PYTHIA6  \cite{Sjostrand:2006za} for parton shower, hadronisation and heavy flavour decays while PGS \cite{PGS} was the detector emulator, supplemented by a generic  LHC parameter card. In particular, the detector parameters simulated were as follows.  We  considered a calorimeter coverage $|\eta|<5.0$, with segmentation $\Delta\eta\times\Delta\phi=0.087\times 0.10$ (the number of division in $\eta$ and $\phi$ were 320 and 200, respectively). Moreover, we used Gaussian energy resolution, with
\begin{equation}
\frac{\Delta E}{E}=\frac{a}{\sqrt{E}}\oplus b,
\end{equation}
 where $a=0.5$ and $b=0.03$ for both the Electro-Magnetic (EM) and hadron calorimeter resolution, with $\oplus $ meaning addition in  quadrature. The algorithm to perform jet finding was a ``cone" one with jet radius $\Delta R=0.5$. The calorimeter trigger cluster finding a seed(shoulder) threshold was $5$ GeV($1$ GeV). 
Further, the kinematic behaviour of the final state particles was mapped with the help of  MadAnalysis5 \cite{Conte:2012fm}. 

For  the MC analysis, six masses were selected for the charged Higgs boson: $120$, $170$, $200$ $400$, $500$ and $750$ GeV. For each such values, the dominant background is the irreducible one induced by $q\bar q'\rightarrow W^{\pm} \rightarrow \tau\bar \nu_{\tau}$, even if $M_{H^\pm}$ is always significantly larger than $M_{W^\pm}$ (indeed, in line with the findings of Ref. \cite{HernandezSanchez:2012eg}). The aforementioned reducible backgrounds, 
  $g {q}'  \to W^{\pm} q$ 
and $q \bar{q}  \to W^{+} W^{-} \to l^{+}l^{-}\nu \nu$, are  smaller in comparison. However, all of these are included in our analysis. 
  As previously stated, we will be looking for leptonic $\tau$ decays, so that the final state is   $l+E_T \hspace{-.4 cm} / $\hspace{.2 cm}, where $l=e,\mu$ and $E_T \hspace{-.4 cm} / $\hspace{.2 cm} is the missing transverse energy. We placed no cuts on the latter while for both lepton and jets {{the following acceptance region in transverse momentum and rapidity was adopted: $p_T(l),\, p_T(j)>10$ GeV and $|\eta(l)|, |\eta(j)|<3$}} with $\Delta R (j,l) >0.5$. In fact, owing to QCD Initial State Radiation (ISR), there could be any number of jets in the final state, however, in our analysis, we will finally select events with at least one lepton  and no jets.

\begin{figure}
\includegraphics[scale=0.4]{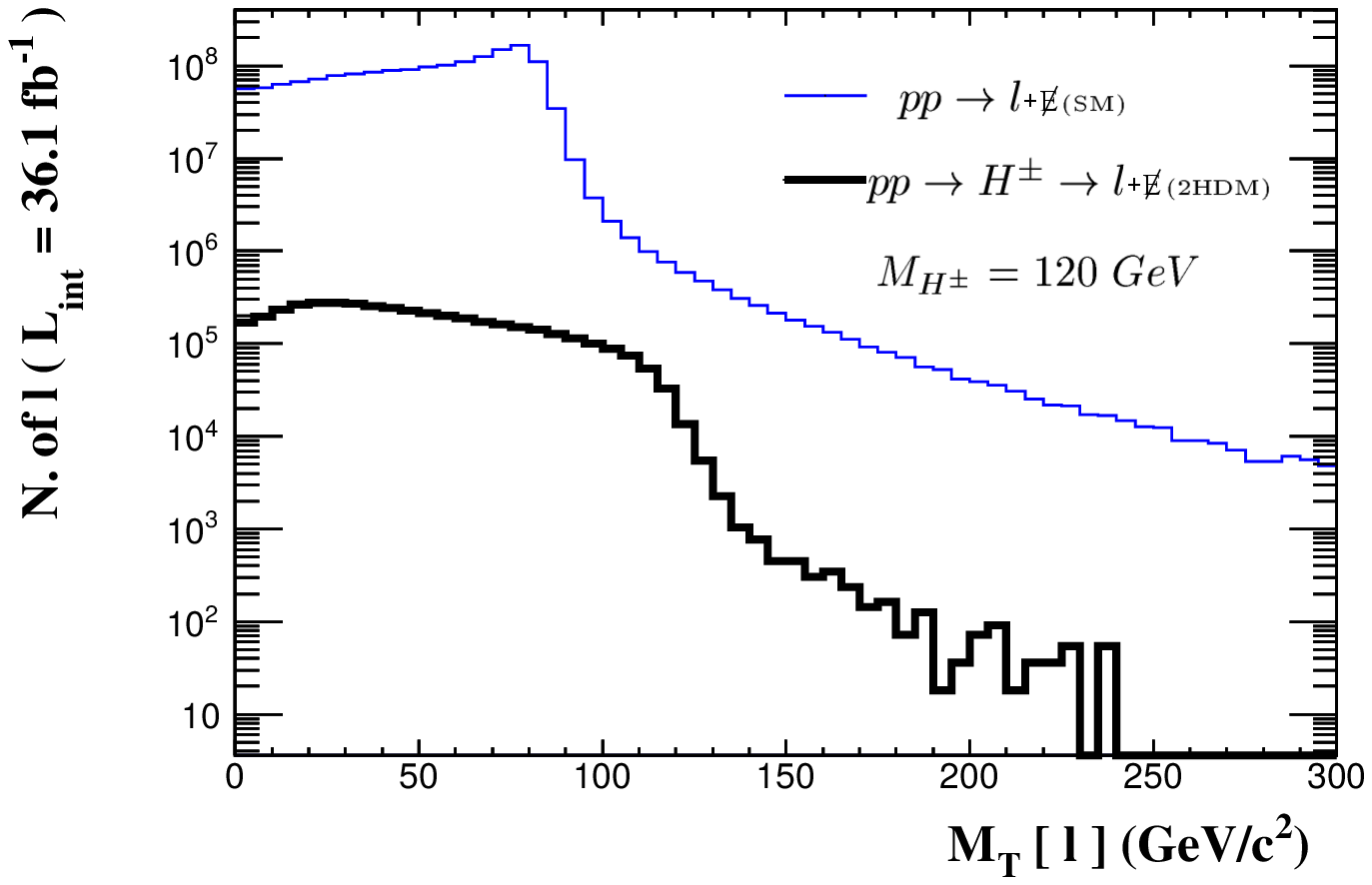}
\includegraphics[scale=0.4]{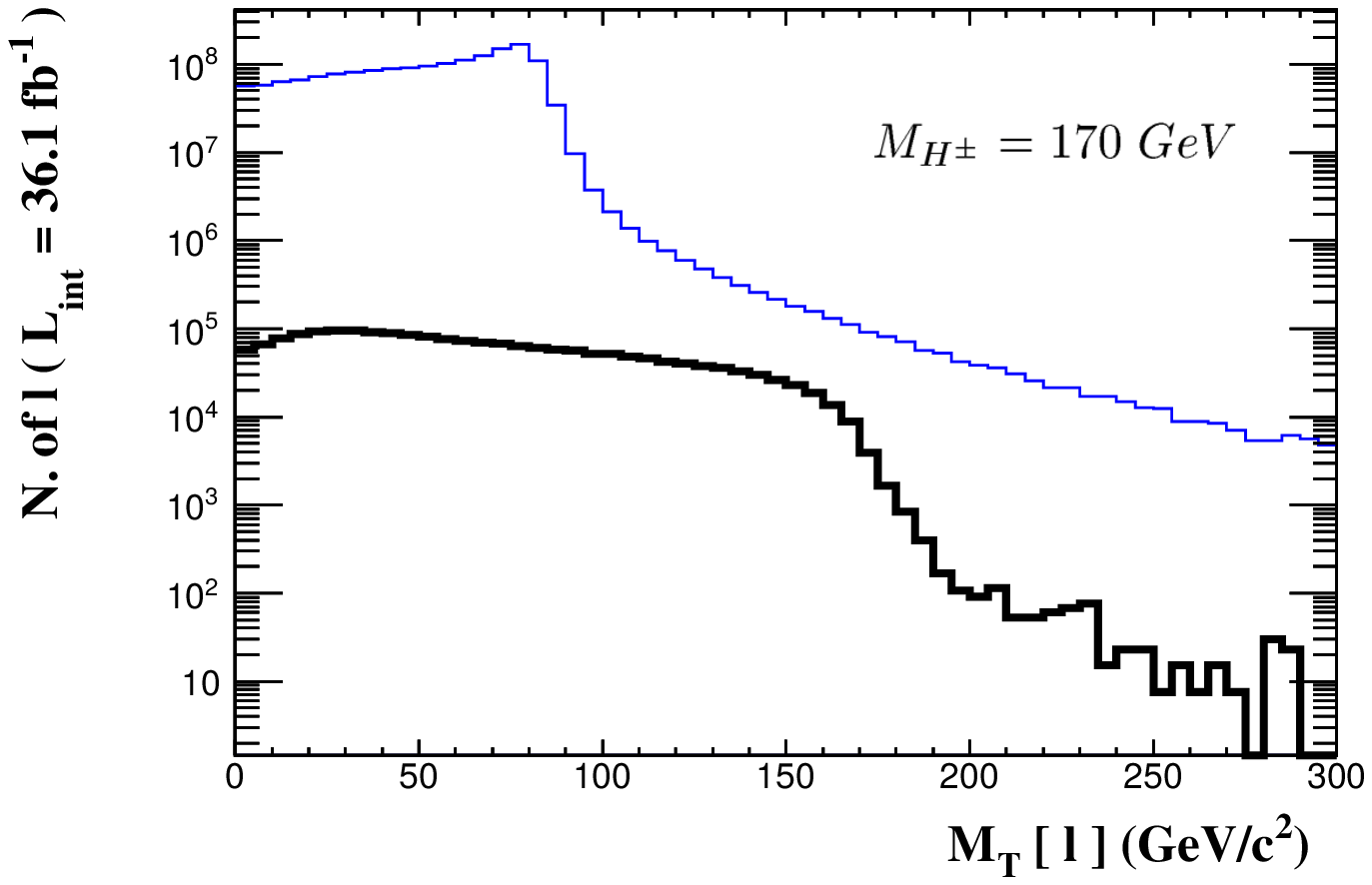}
\caption{Transverse mass plots for signal and background, for selected $M_{H^\pm}$ choices of the former. No cuts are here applied.}
\label{pptaunu-cut-0}
\end{figure}

Since the invariant mass of the final state is not reconstructible, as previously  done  \cite{HernandezSanchez:2012eg}, one can analyse the transverse mass
$M_T(l) \equiv \sqrt{ (E_l^T -E_T \hspace{-.4 cm} / \hspace{.2 cm} )^2 - (p_l^x+p_{miss}^x)^2 - (p_l^y+p_{miss}^y)^2 }$, where $p^x_{l,miss}$ and $p^y_{l,miss}$ are located  in the transverse plane, thus assuming that the proton beams are along the $z$-axis.
In Fig. \ref{pptaunu-cut-0}, the shape of the transverse mass is reconstructed at detector level without selection cuts,  wherein both signal and background can be seen, which reinforces the fact that,  at the differential level (e.g., for $m_{H^\pm}=120$ and 170 GeV, although the situation is the same for any other mass), the potential Jacobian peak correlating to the charged Higgs boson mass is well beyond the background distribution. Hence, a careful signal selection will be proposed which preserves such a difference as much as possible. In particular, we will optimise this to the given value of the charged Higgs boson  mass. That is,  a trial and error approach will be assumed, wherein the $M_{H^\pm}$ value is an input parameter to the kinematic analysis and the selection cuts adopted depend on it.

In order to fully define our selection, let us now investigate some relevant differential distributions that can be used to enhance the signal-to-background rate. (Notice that, for reasons of space, we will not show all charged Higgs mass values in each case.)

\begin{figure}
\includegraphics[scale=0.4]{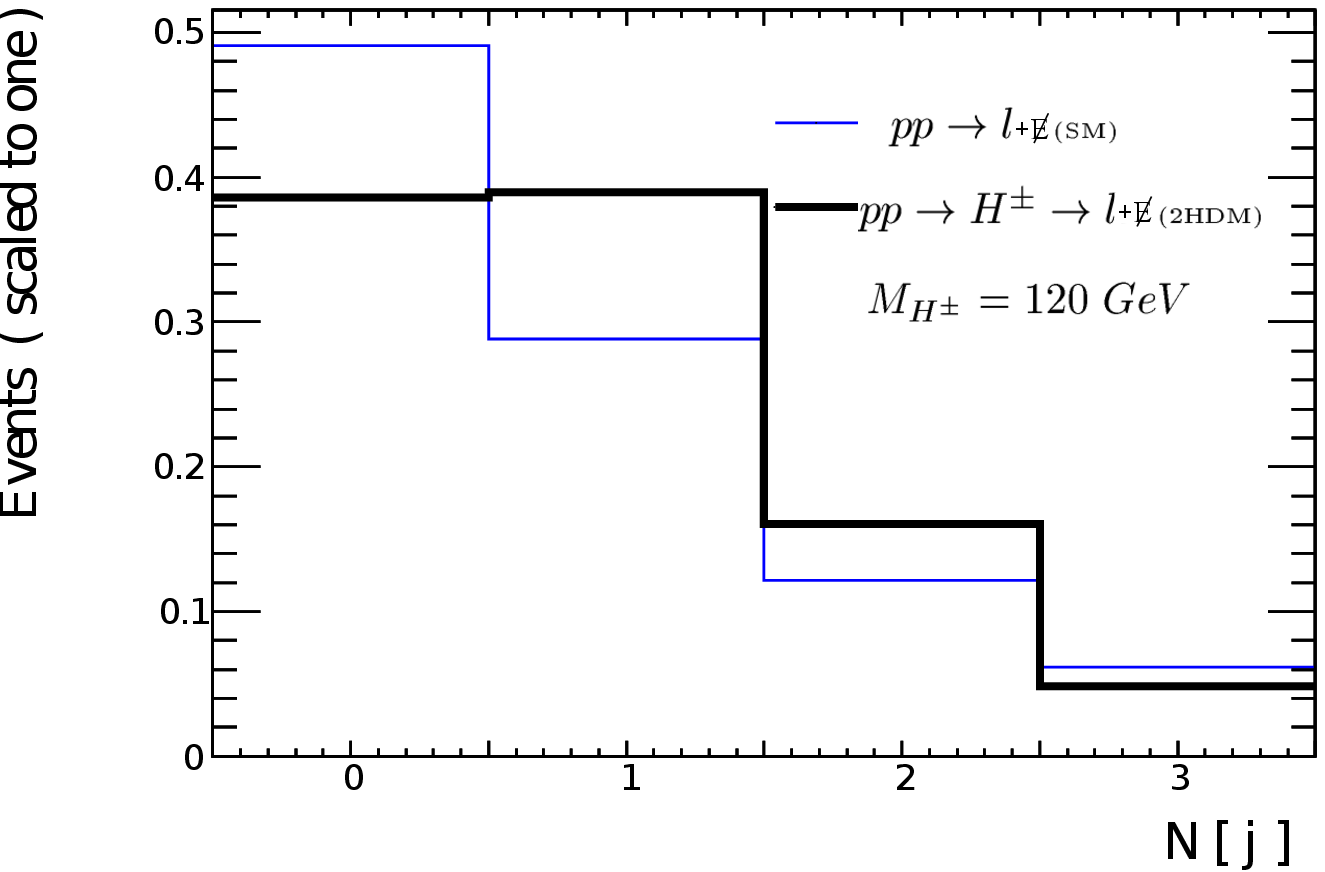}
\includegraphics[scale=0.4]{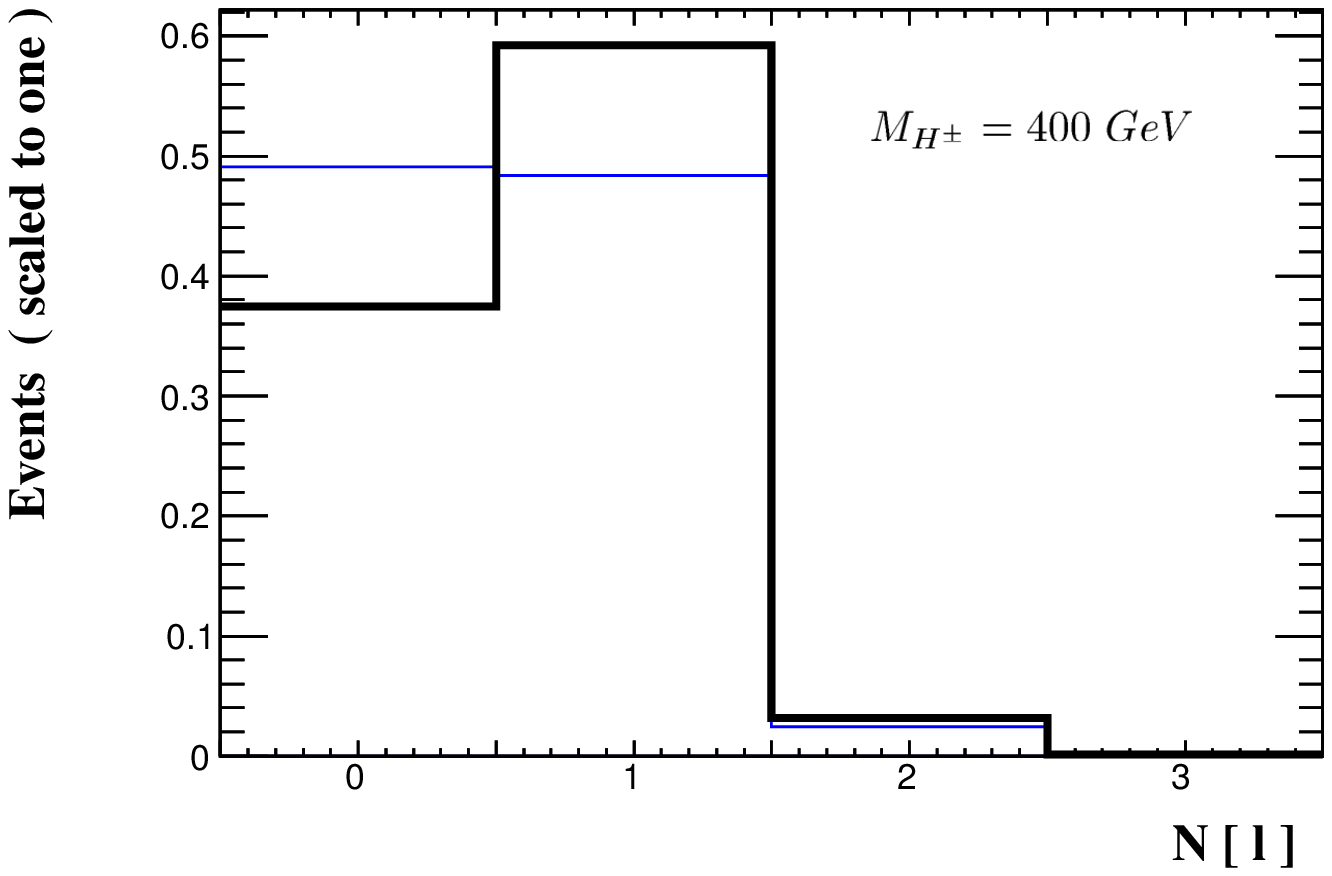}
\includegraphics[scale=0.4]{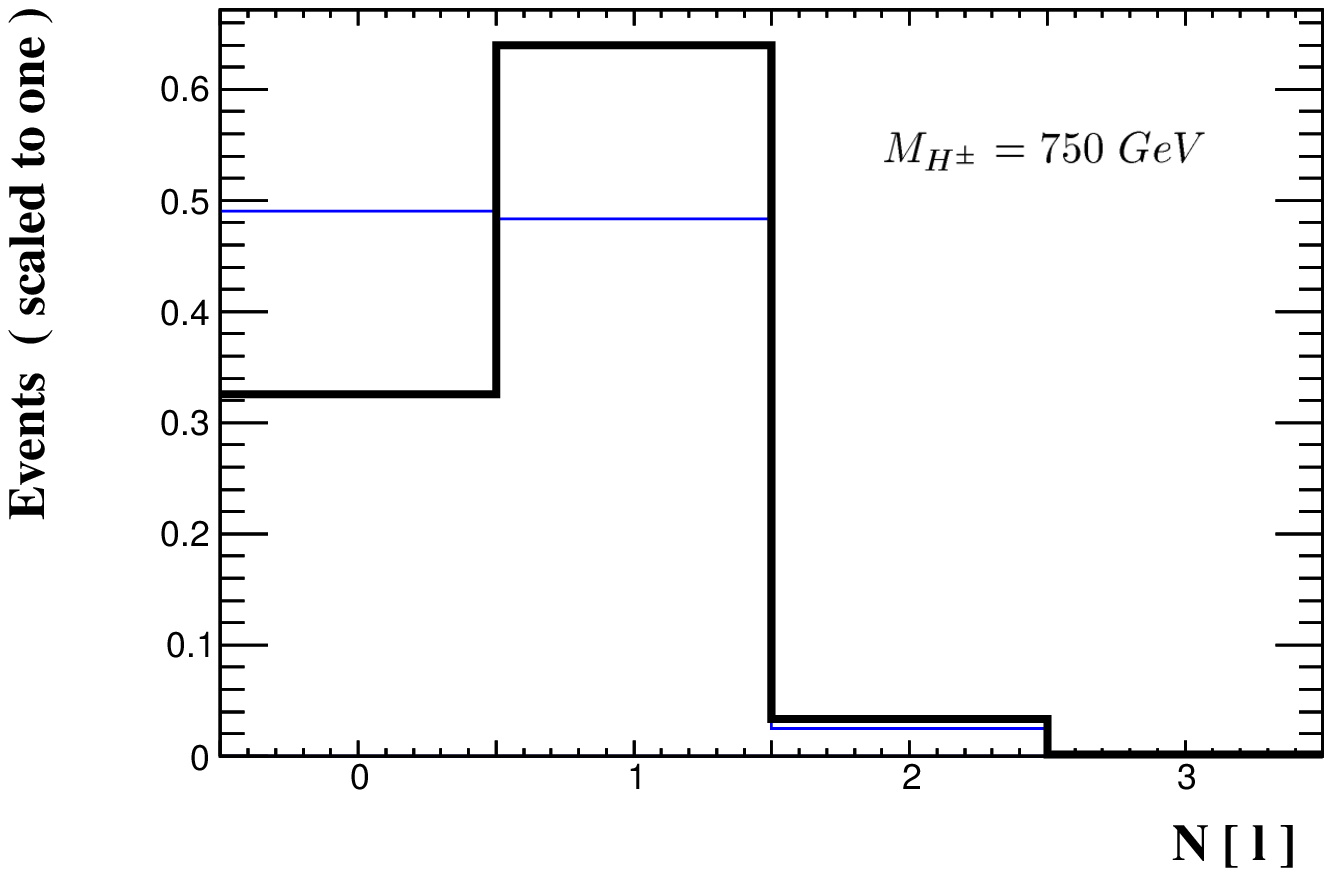}
\caption{Lepton multiplicity plots for signal and background, for selected $M_{H^\pm}$ choices of the former, over the acceptance region for leptons and jets, in both transverse momentum as pseudorapidity.}
\label{pptaunu-cut-a}
\end{figure}
\begin{figure}
\includegraphics[scale=0.4]{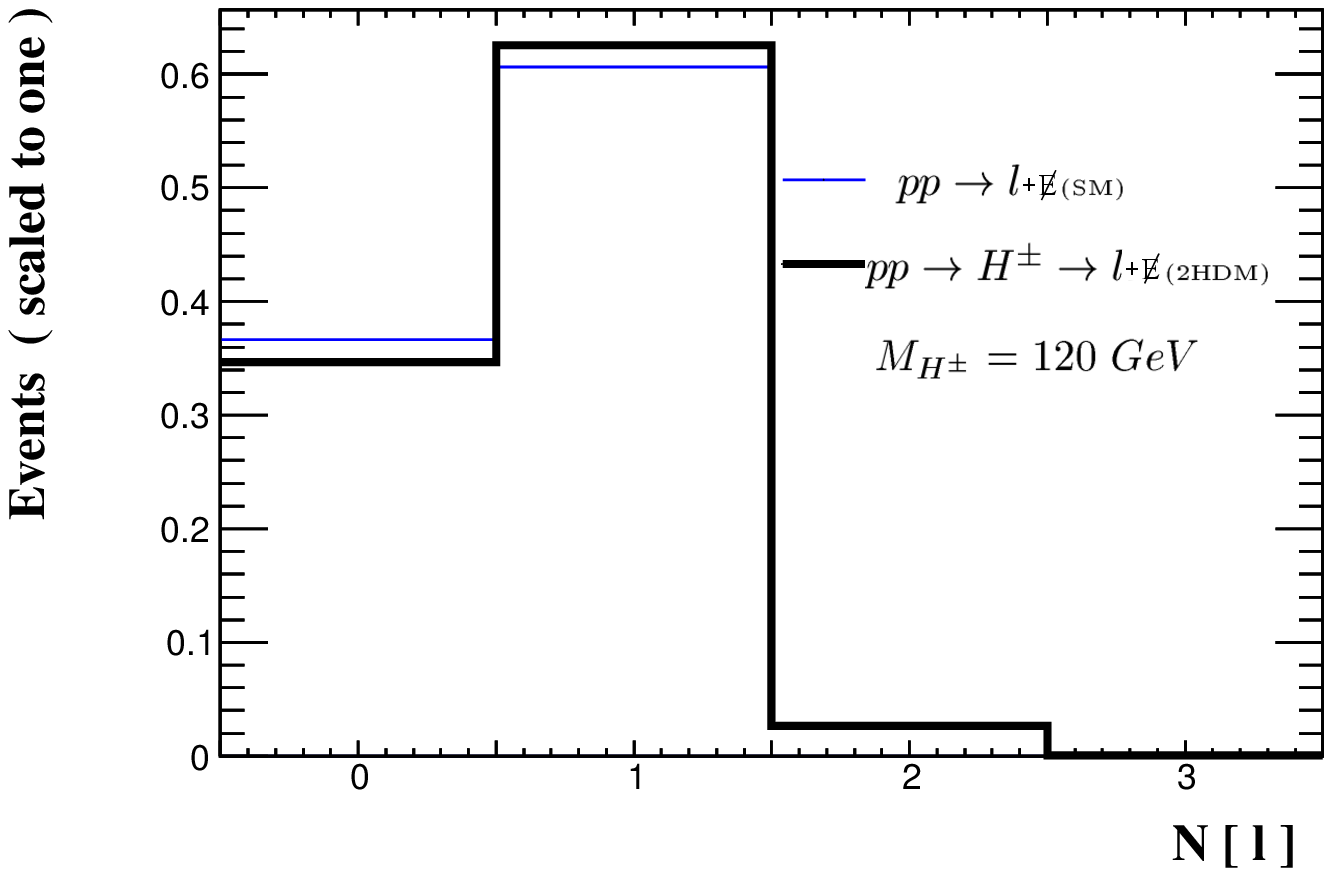}
\includegraphics[scale=0.4]{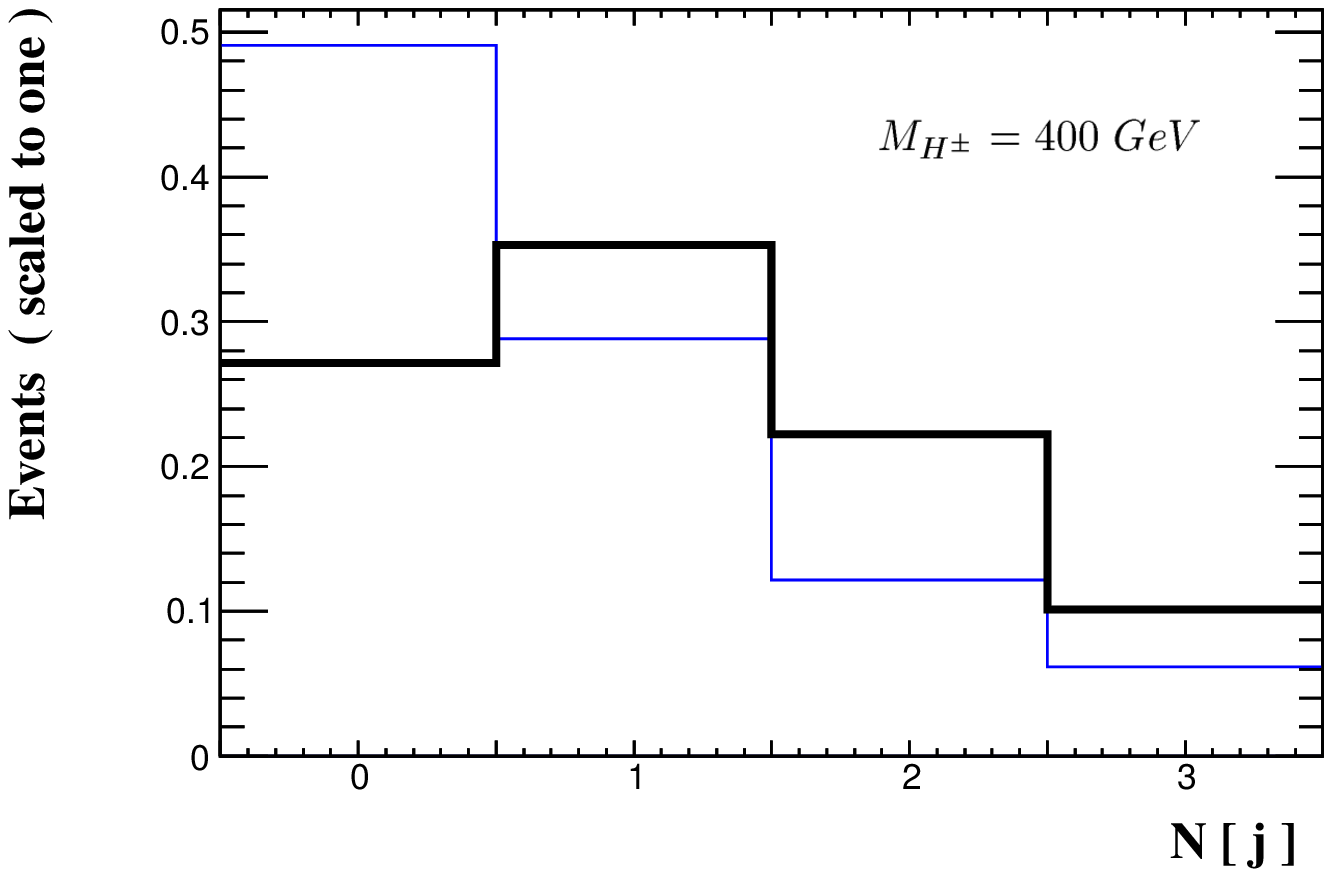}
\includegraphics[scale=0.4]{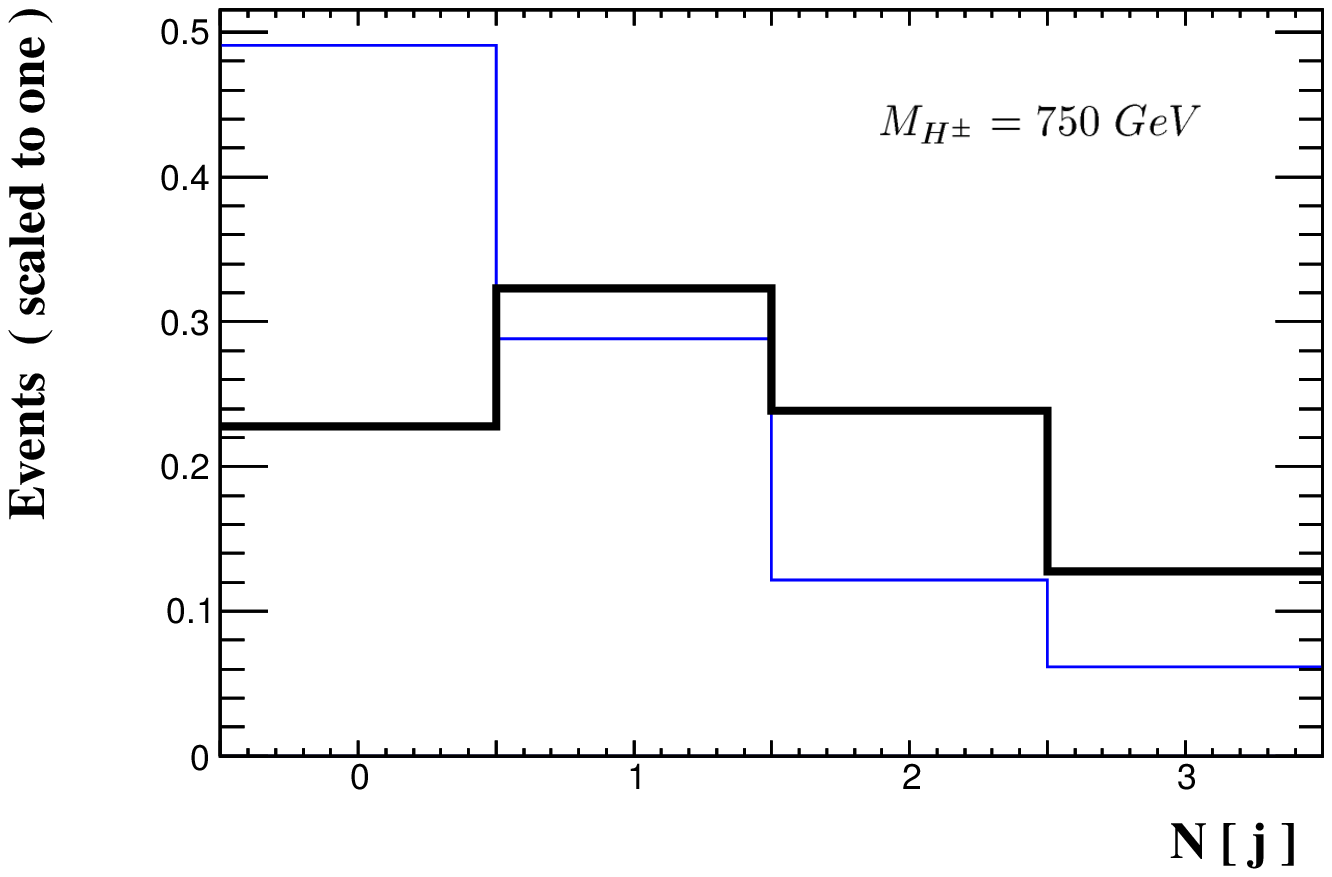}
\caption{Hadron multiplicity plots for signal and background, for selected $M_{H^\pm}$ choices of the former, over the acceptance region for leptons and jets, in both transverse momentum and pseudorapidity.}
\label{pptaunu-cut-b}
\end{figure}
\begin{figure}
\includegraphics[scale=0.4]{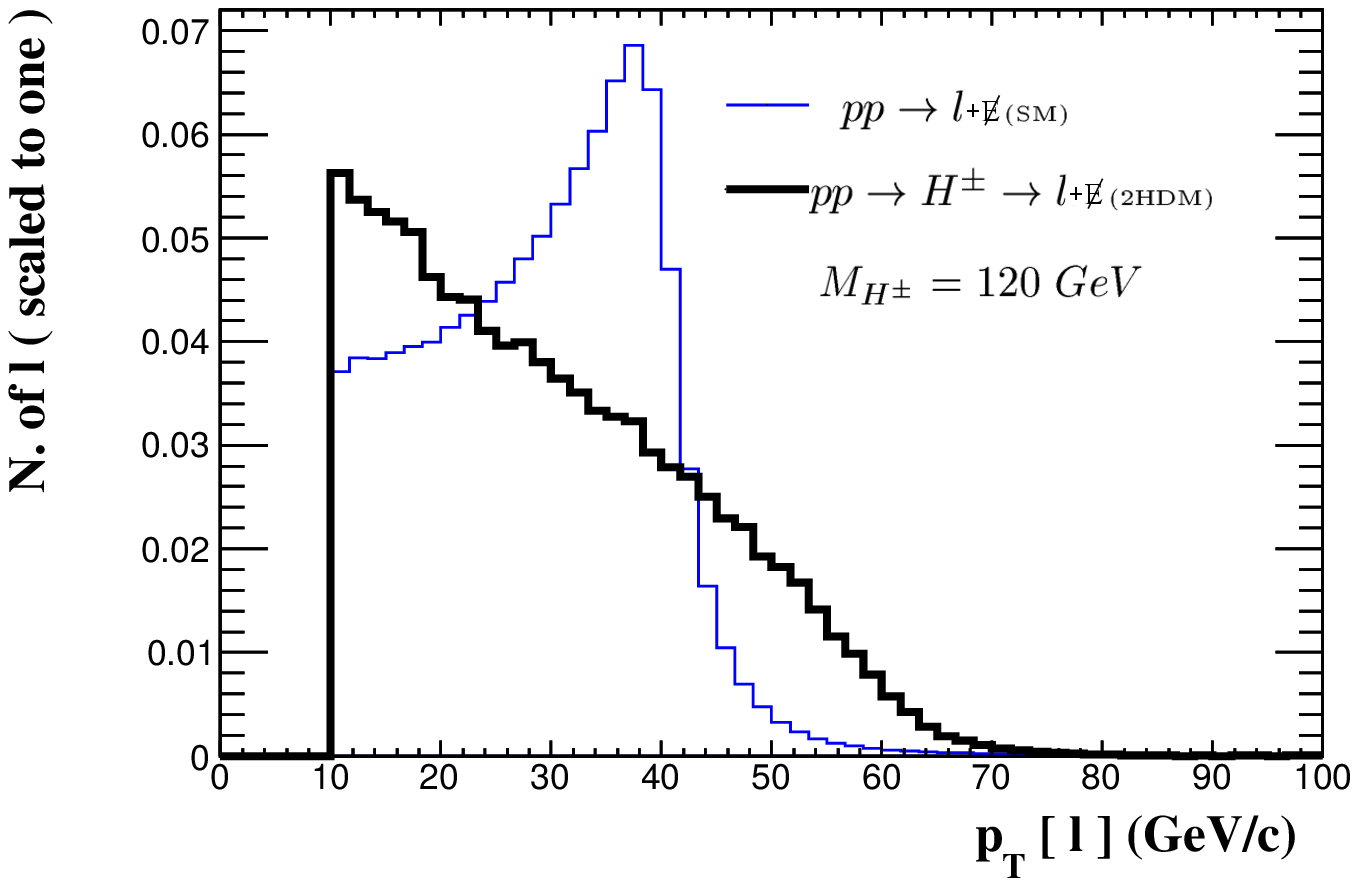}
\includegraphics[scale=0.4]{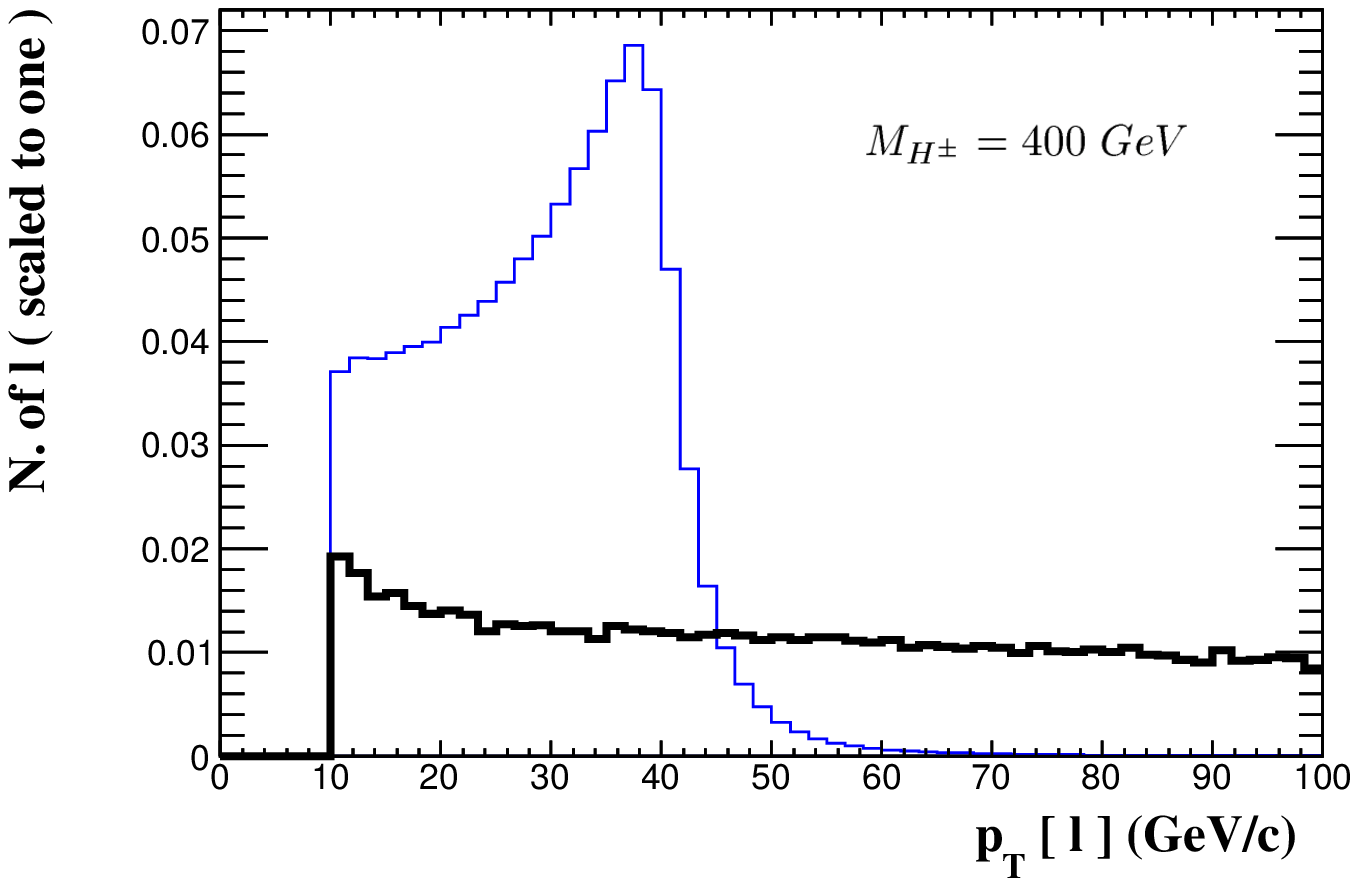}
\includegraphics[scale=0.4]{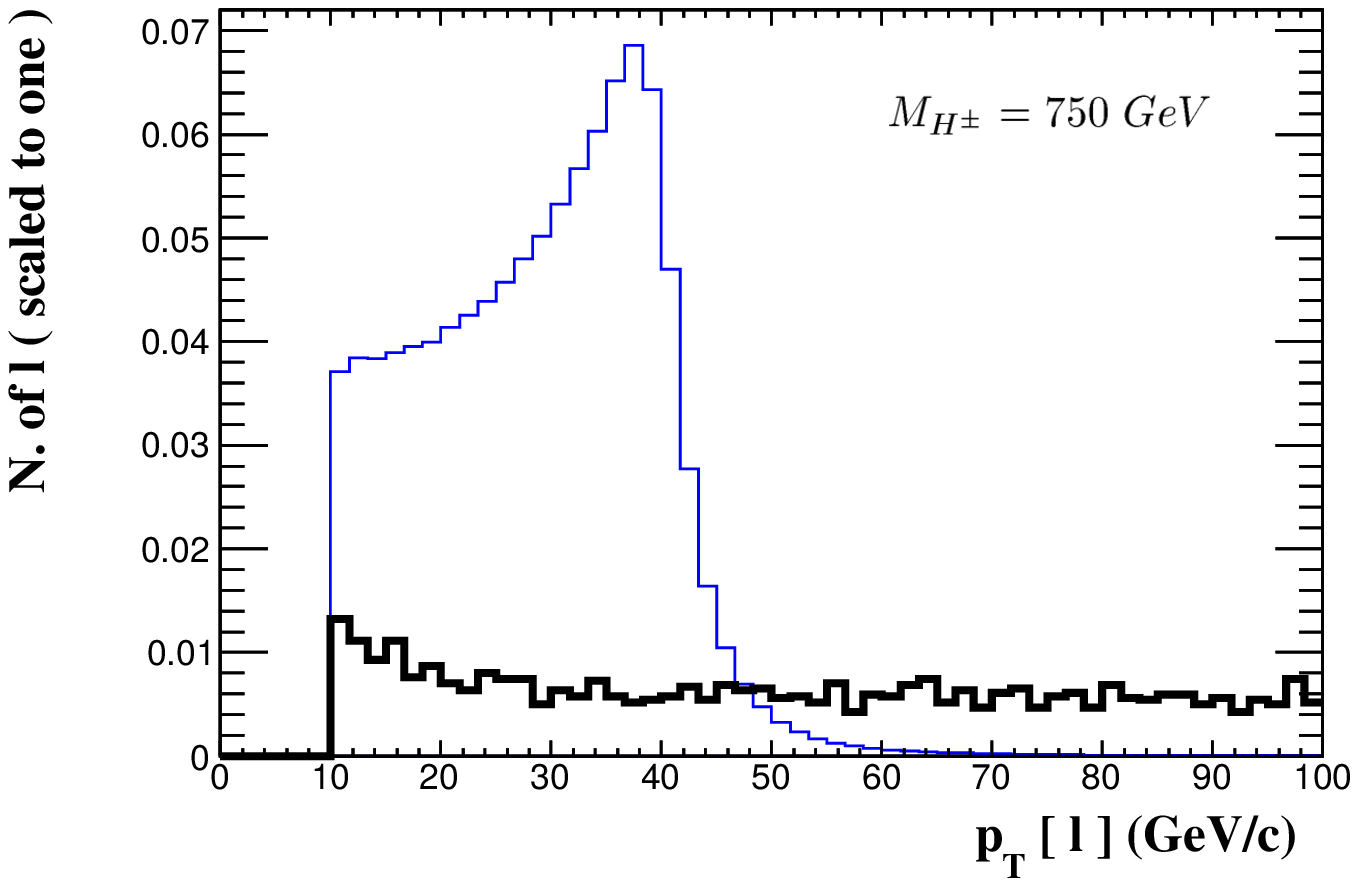}
\caption{Leptonic transverse momentum plots for signal and background, for selected $M_{H^\pm}$ choices of the former, over the acceptance region for leptons and jets, in both transverse momentum as pseudorapidity. Further, jets are vetoed here.}
\label{pptaunu-cut-1}
\end{figure}
\begin{figure}
\includegraphics[scale=0.4]{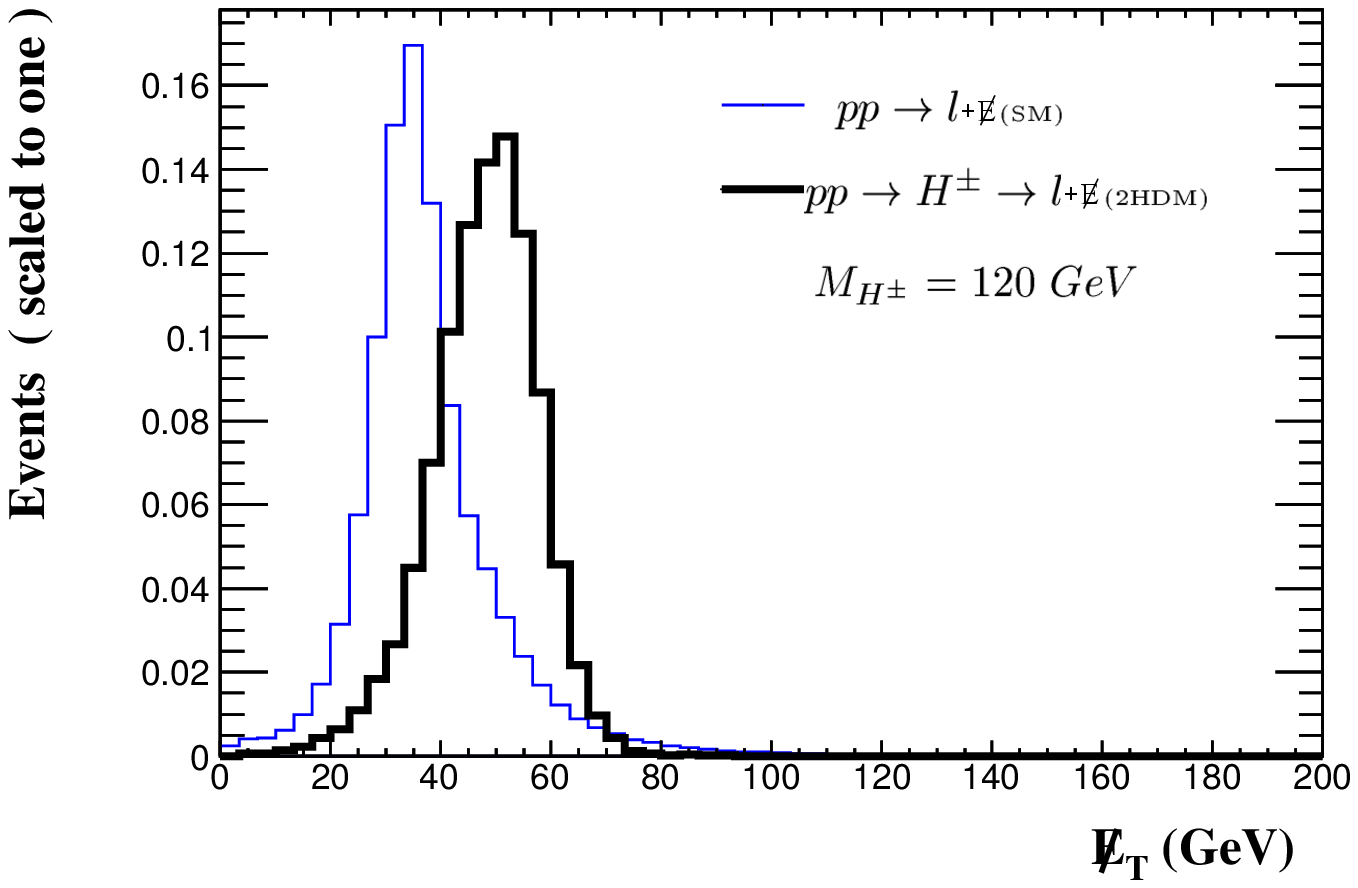}
\includegraphics[scale=0.4]{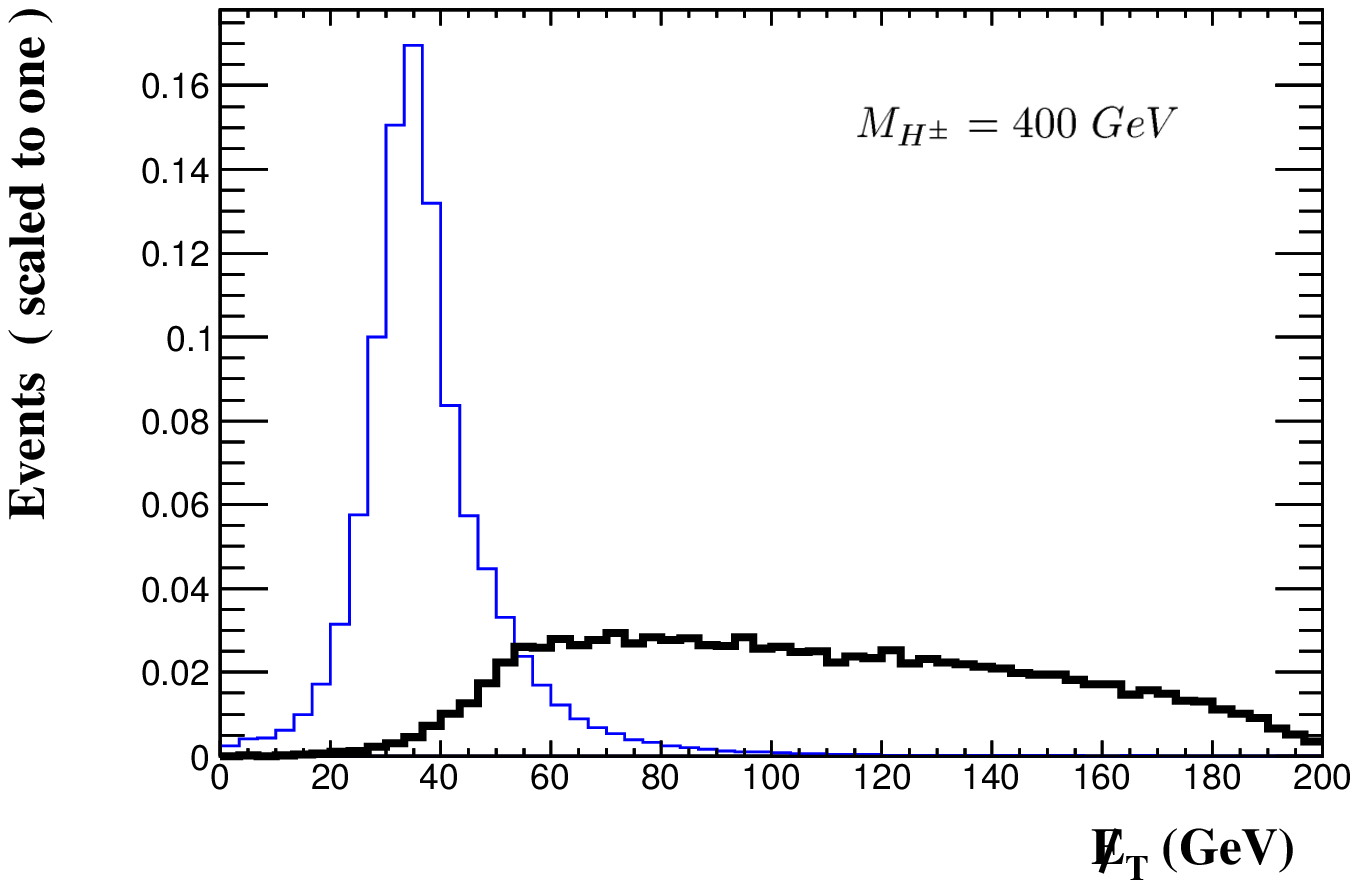}
\includegraphics[scale=0.4]{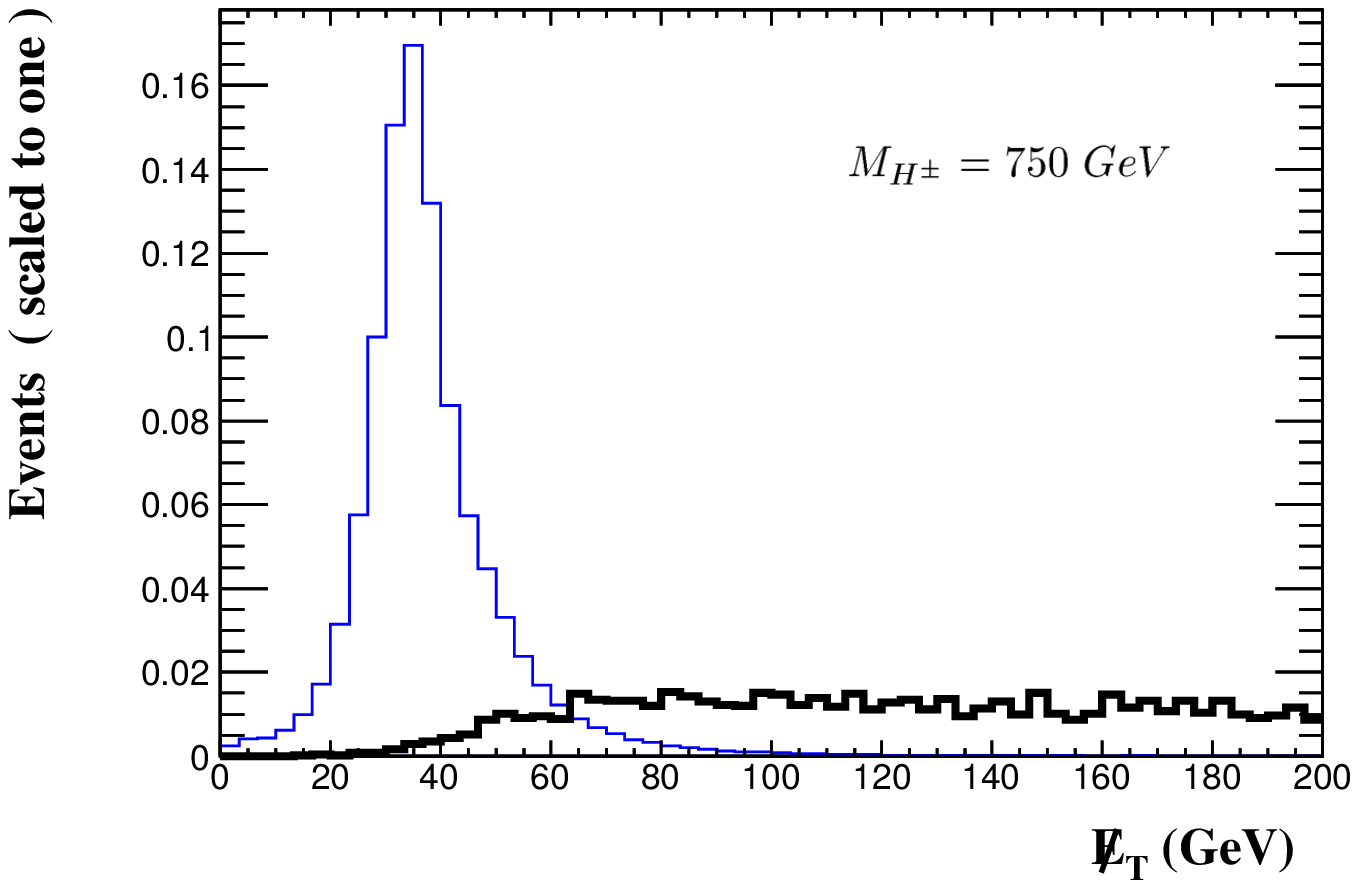}
\caption{Missing transverse energy plots  for signal and background, for selected $M_{H^\pm}$ choices of the former, over the acceptance region for leptons and jets, in both transverse momentum as pseudorapidity. Further, jets are vetoed here.}
\label{pptaunu-cut-2}
\end{figure}
\begin{figure}
\includegraphics[scale=0.4]{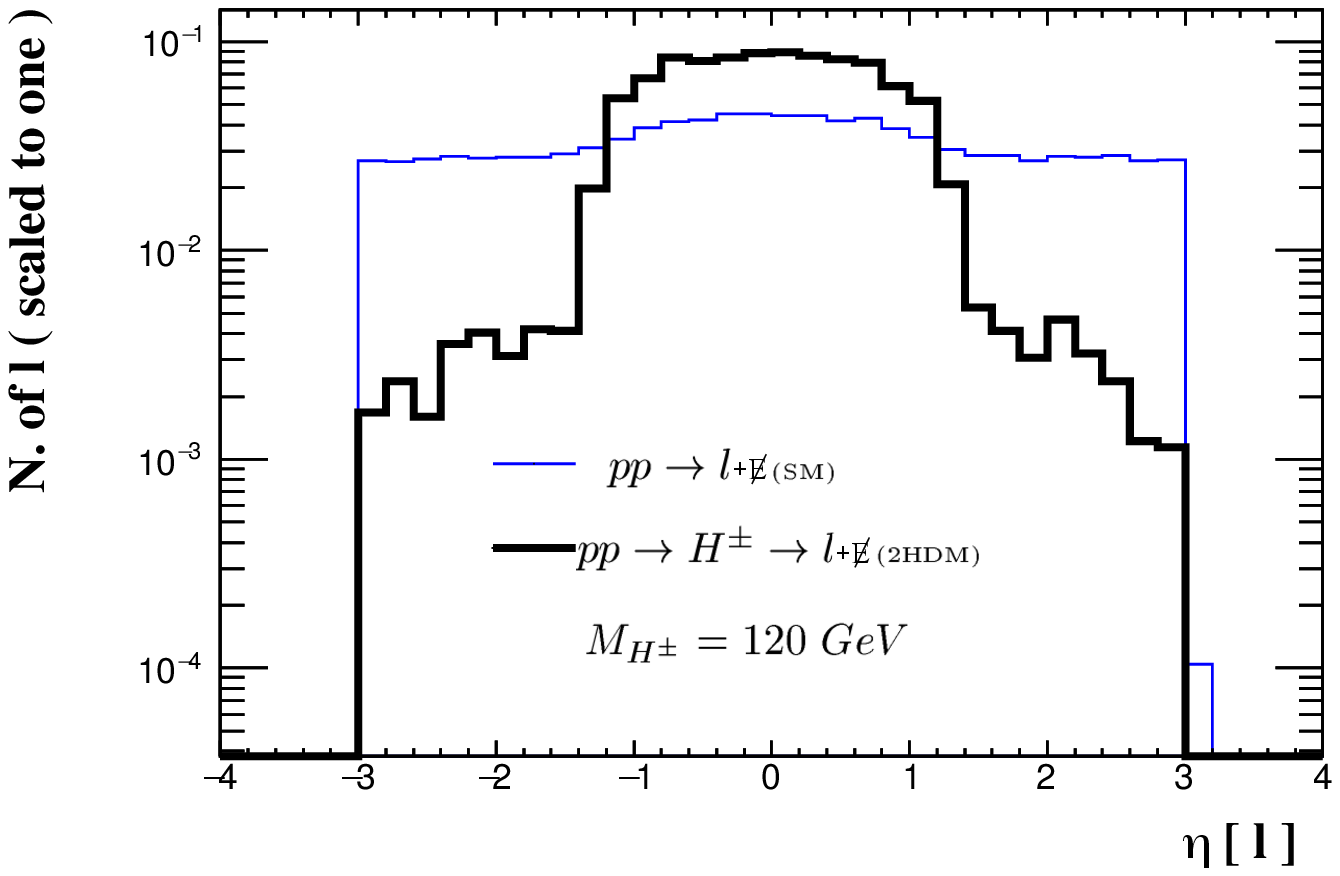}
\includegraphics[scale=0.4]{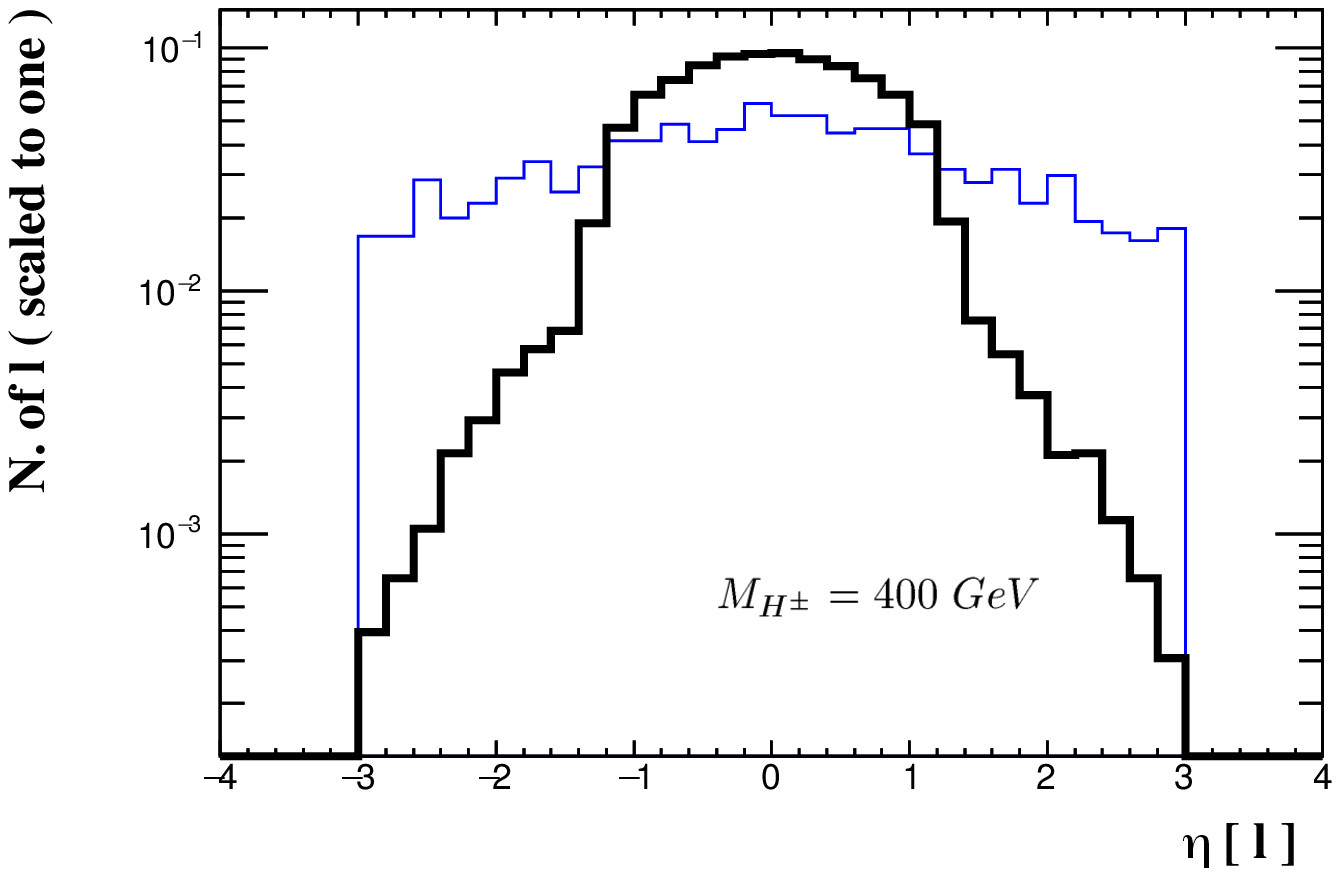}
\includegraphics[scale=0.4]{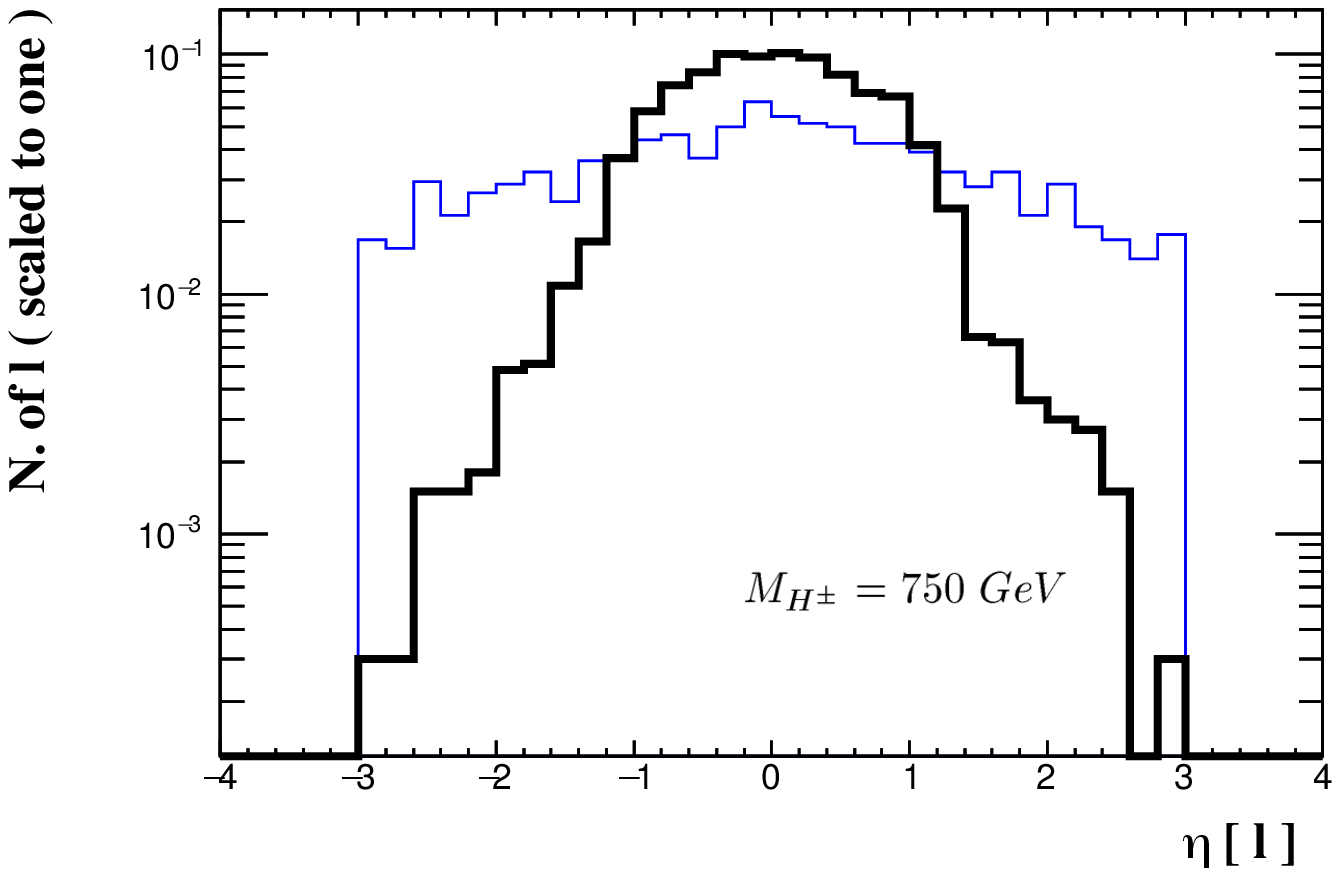}
\caption{Pseudorapidity plots  for signal and background, for selected $M_{H^\pm}$ choices of the former, over the acceptance region for leptons and jets, in both transverse momentum as pseudorapidity. Further, jets are vetoed here.}
\label{pptaunu-cut-3}
\end{figure}
\begin{figure}
\includegraphics[scale=0.4]{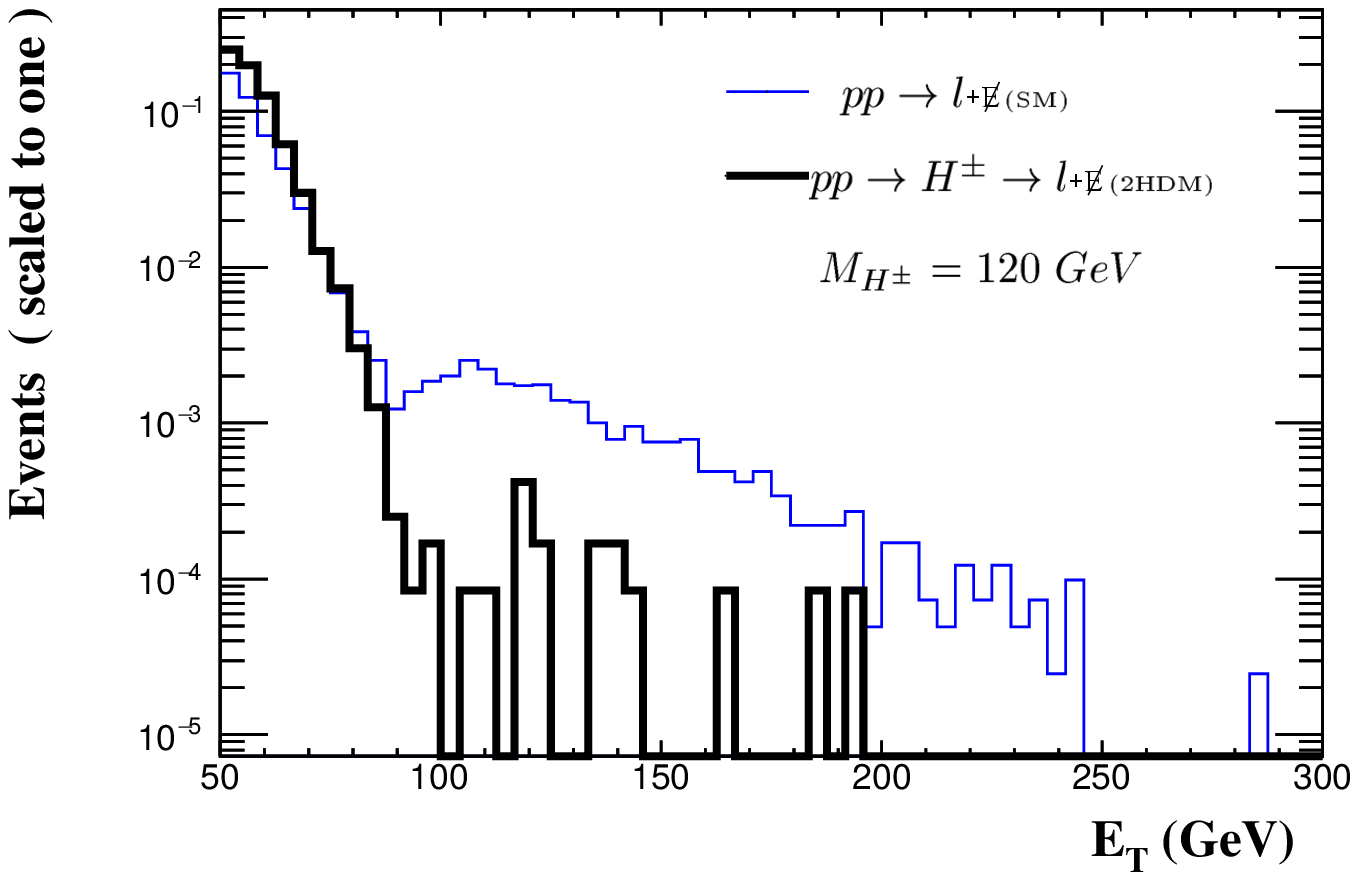}
\includegraphics[scale=0.4]{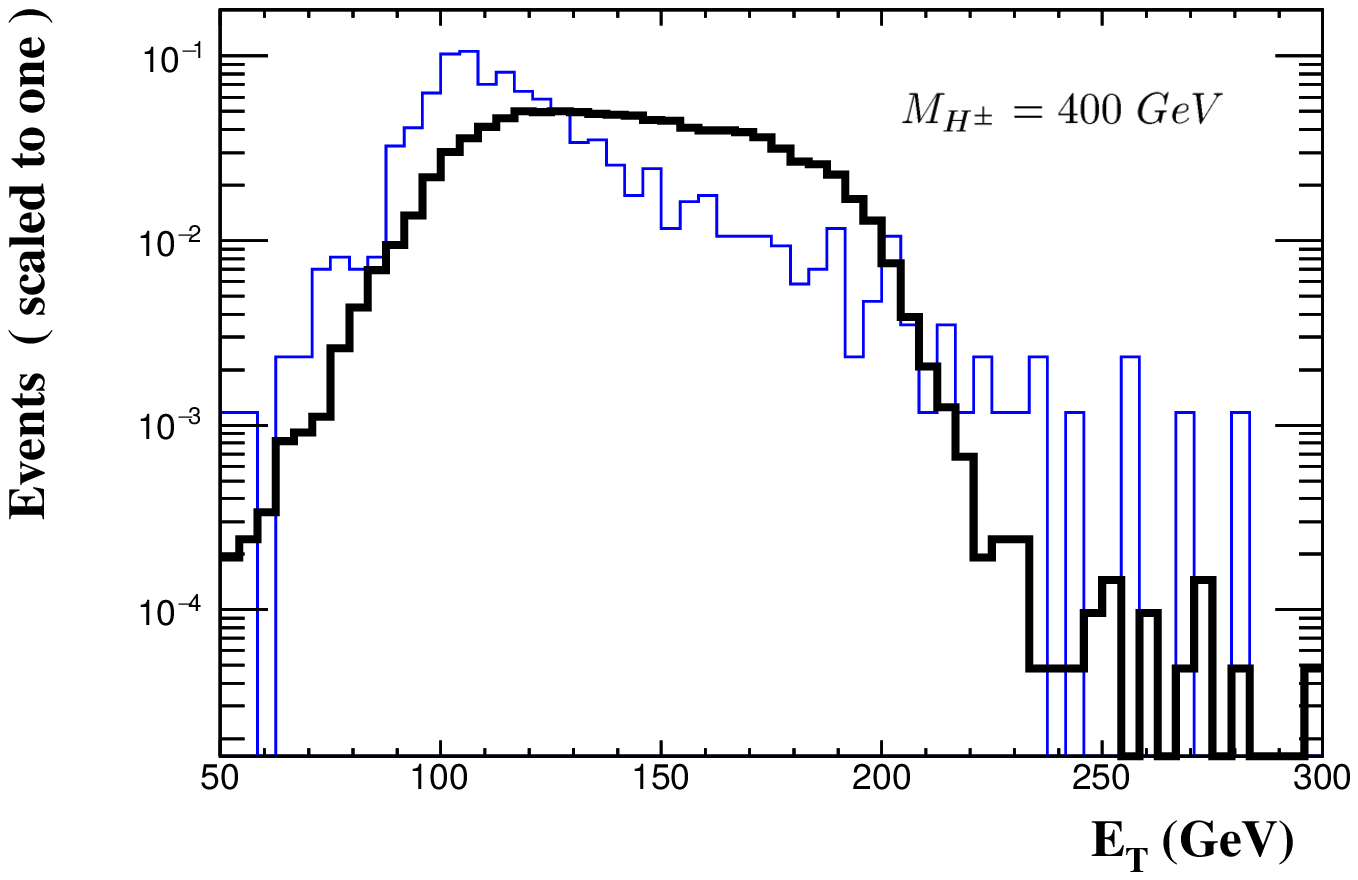}
\includegraphics[scale=0.4]{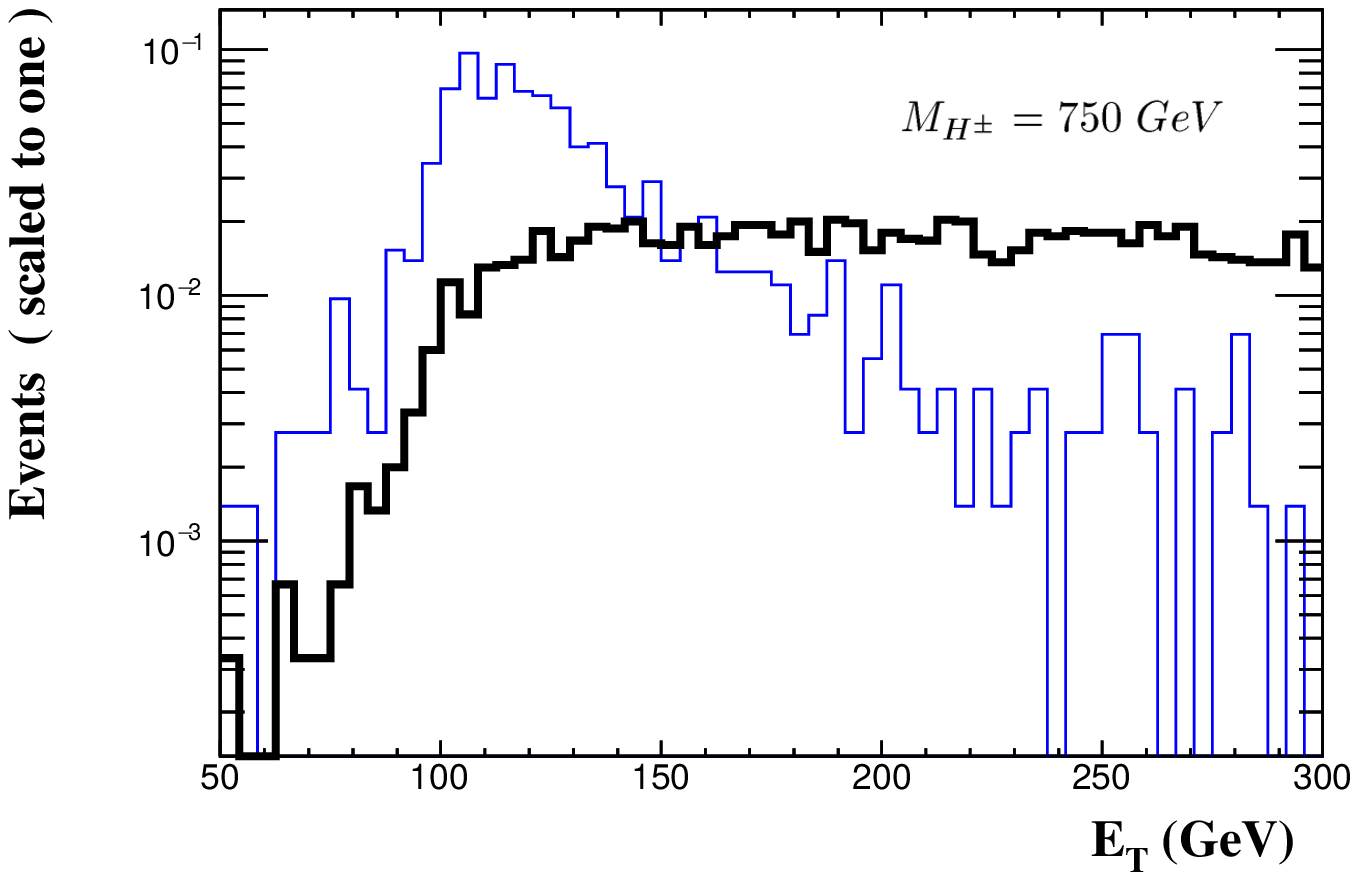}
\caption{Missing transverse energy plots  for signal and background, for selected $M_{H^\pm}$ choices of the former, over the acceptance region for leptons and jets, in both transverse momentum as pseudorapidity. Further, jets are vetoed here.}
\label{pptaunu-cut-4}
\end{figure}
\begin{figure}
\includegraphics[scale=0.4]{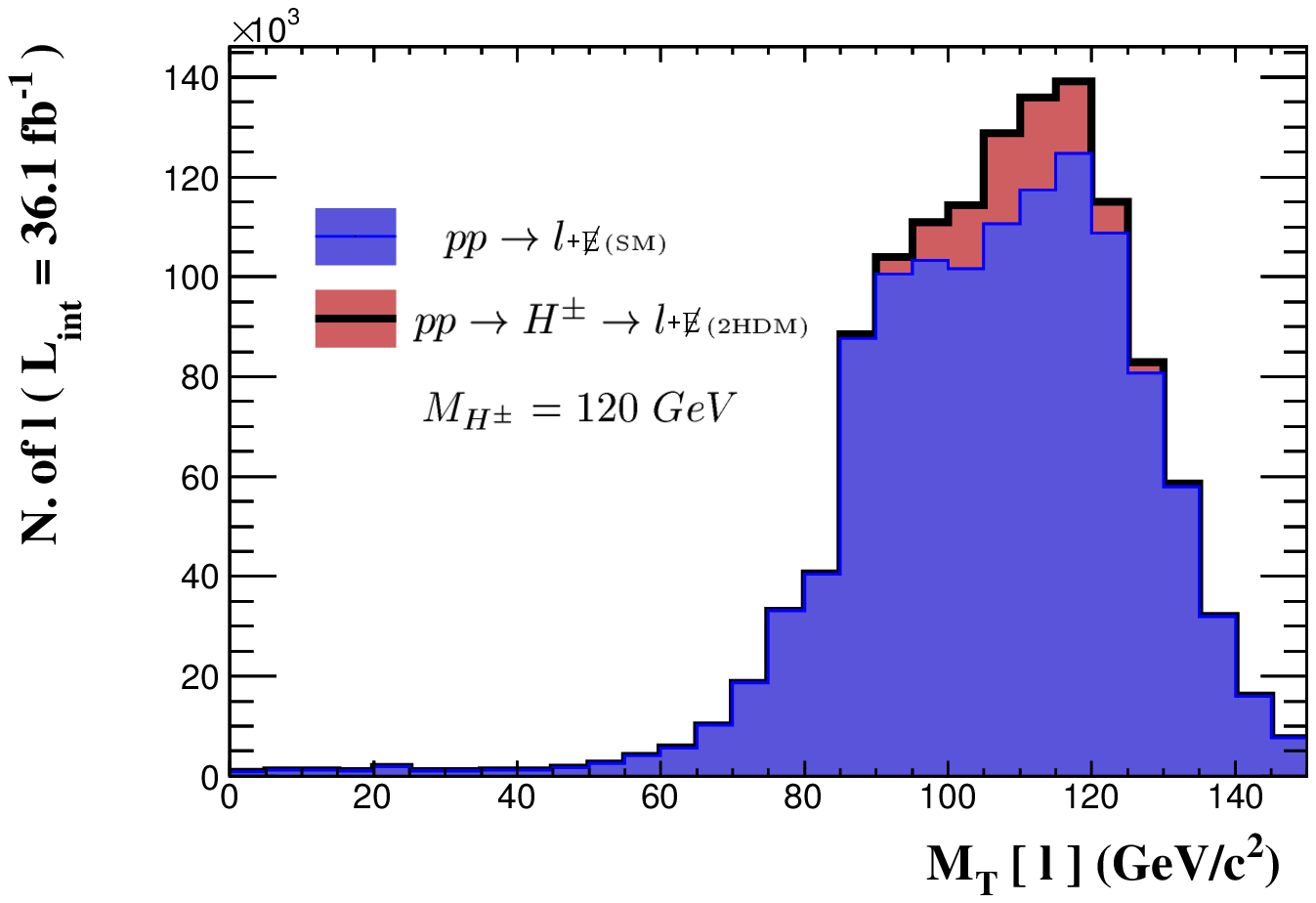}
\includegraphics[scale=0.4]{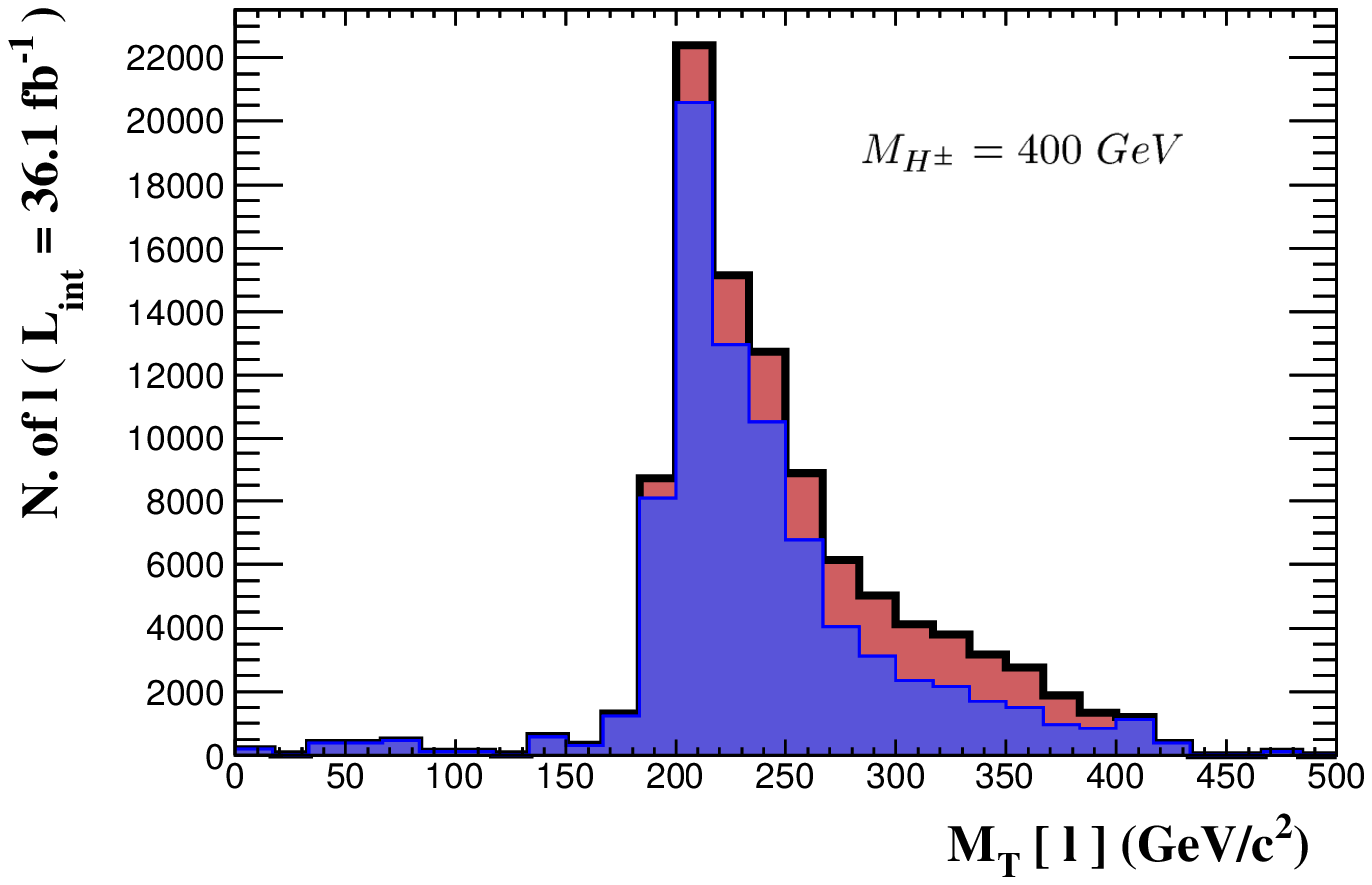}
\includegraphics[scale=0.4]{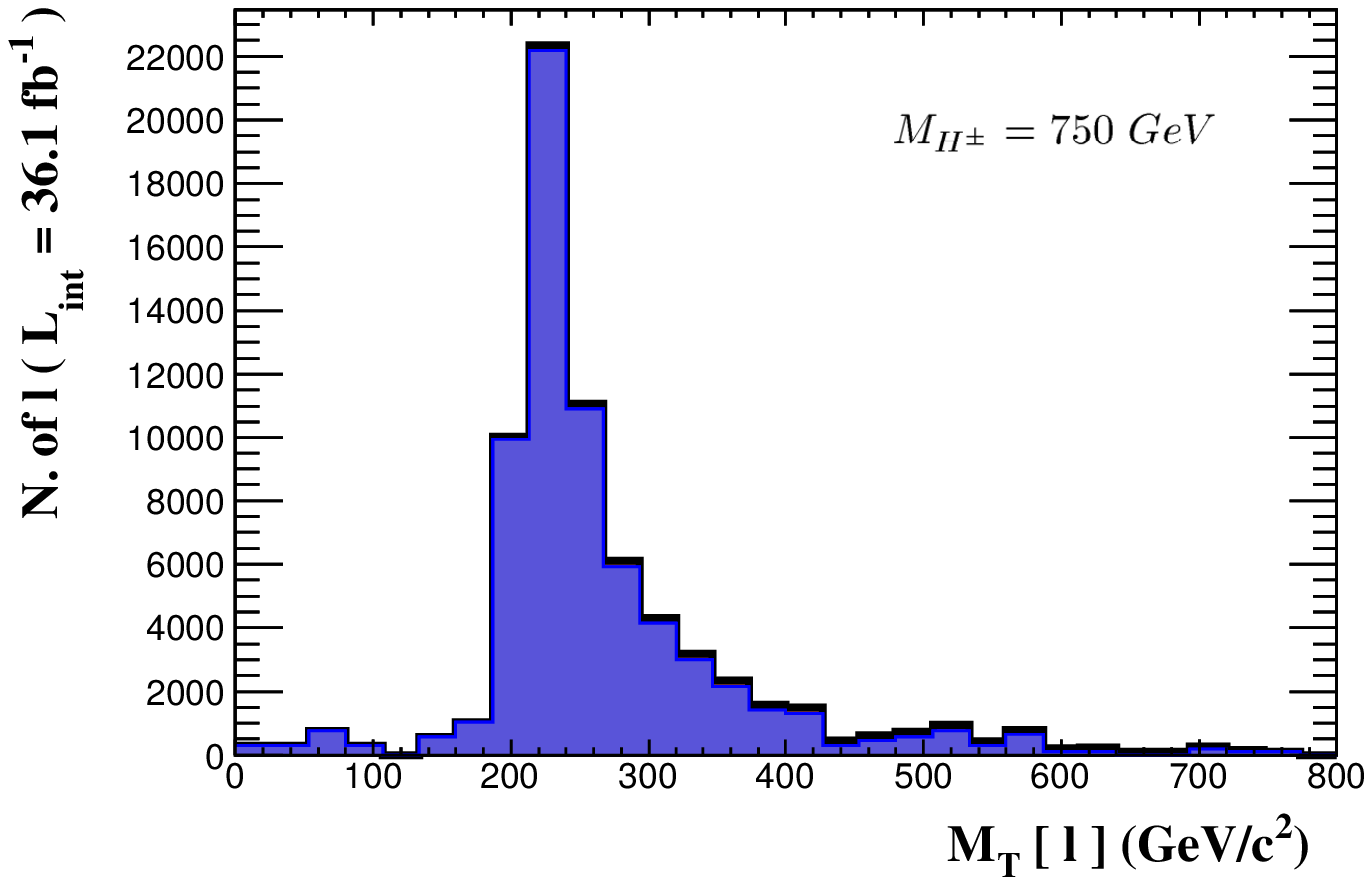}
\caption{Transverse mass plots for signal and background, for selected $M_{H^\pm}$ choices of the former, after Cuts 1--4. Histograms are stacked here.}
\label{pptaunu-cut-7}
\end{figure}

\begin{enumerate}
\item From the lepton and hadronic multiplicity plots, see  Figs. \ref{pptaunu-cut-a} and \ref{pptaunu-cut-b}, we require at least one lepton and impose no jets in our sample. (Here, we impose on  both lepton and jets {{the acceptance region in transverse momentum and rapidity as already discussed: i.e., $p_T(l),\, p_T(j)>10$ GeV and $|\eta(l)|, |\eta(j)|<3$}} with $\Delta R (j,l) >0.5$.) Further, 
 by looking at Fig. \ref{pptaunu-cut-1} (wherein the jet veto is applied), it can be seen that the cut $p_T(l)\geq 45$ GeV on the leptonic transverse momentum can be profitably adopted for all charged Higgs mass boson masses. 
\item The missing transverse energy plots, Fig. \ref{pptaunu-cut-2}, suggest the use of the following cuts: for $M_{H^{\pm}}=120$ GeV, $40$ GeV $\leq \slashed{E}_T \leq 70$ GeV; for $M_{H^{\pm}}=170$ GeV, $60$ GeV $\leq \slashed{E}_T \leq 90$ GeV; for $M_{H^{\pm}}=200$ GeV, $70$ GeV $\leq \slashed{E}_T \leq 105$ GeV; for $M_{H^{\pm}}=400$ GeV, $100$ GeV $\leq \slashed{E}_T \leq 225$ GeV; for $M_{H^{\pm}}=500$ GeV, $90$ GeV $\leq \slashed{E}_T \leq 270$ GeV; for $M_{H^{\pm}}=750$ GeV, $105$ GeV $\leq \slashed{E}_T$.
\item The lepton  pseudorapidity, Fig. \ref{pptaunu-cut-3}, shows that an optimal cut can be defined for all charged Higgs boson  masses as $ | \eta(l)| \leq 1.2$.
\item The total   energy, Fig. \ref{pptaunu-cut-4}, shows that the following cuts can be efficient: 
for $M_{H^{\pm}}=120,\ 170$ GeV, $E_{T} \geq 55$ GeV; for $M_{H^{\pm}}=200$ GeV, $E_{T} \geq 60$ GeV; for $M_{H^{\pm}}=400$ GeV, $E_{T} \geq 80$ GeV; for $M_{H^{\pm}}=500$ GeV, $E_{T} \geq 75$ GeV; for $M_{H^{\pm}}=750$ GeV, $E_{T} \geq 80$ GeV.
\item The transverse mass plots in Fig. \ref{pptaunu-cut-7}  show that the last cuts to  be defined can be as follows: for $M_{H^{\pm}}=120$ GeV, $85$ GeV $\leq M_T(l) \leq 125$ GeV; for $M_{H^{\pm}}=170$ GeV, $90$ GeV $\leq M_T(l) \leq 175$ GeV; for $M_{H^{\pm}}=200$ GeV, $110$ GeV $\leq M_T(l) \leq 205$ GeV; for $M_{H^{\pm}}=400$ GeV, $170$ GeV $\leq M_T(l) \leq 405$ GeV; for $M_{H^{\pm}}=500$ GeV, $200$ GeV $\leq M_T(l) \leq 505$ GeV; for $M_{H^{\pm}}=750$ GeV, $320$ GeV $\leq M_T(l) \leq 755$ GeV.
\end{enumerate}

\begin{figure}
\includegraphics[scale=0.4]{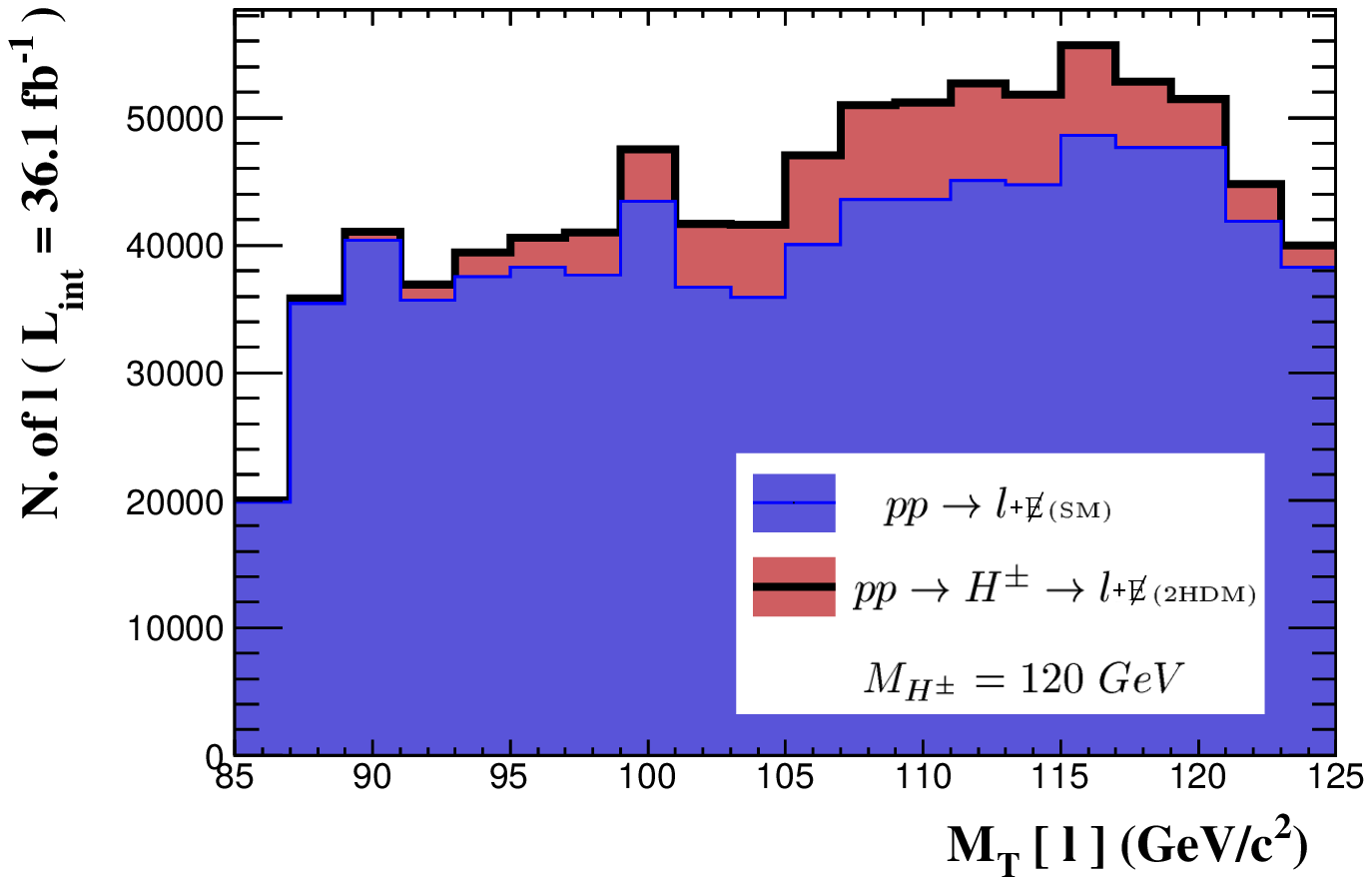}
\includegraphics[scale=0.4]{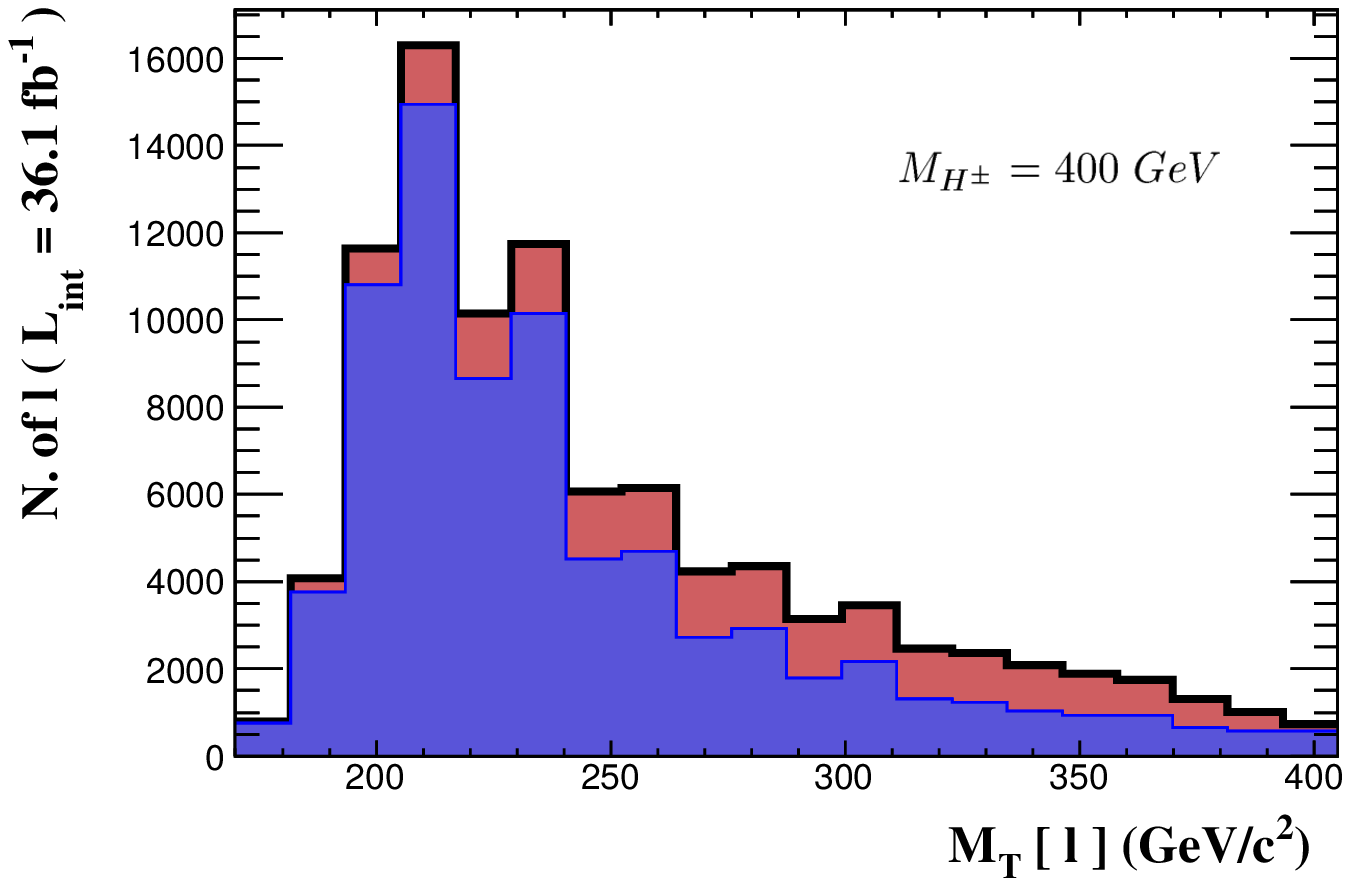}
\includegraphics[scale=0.4]{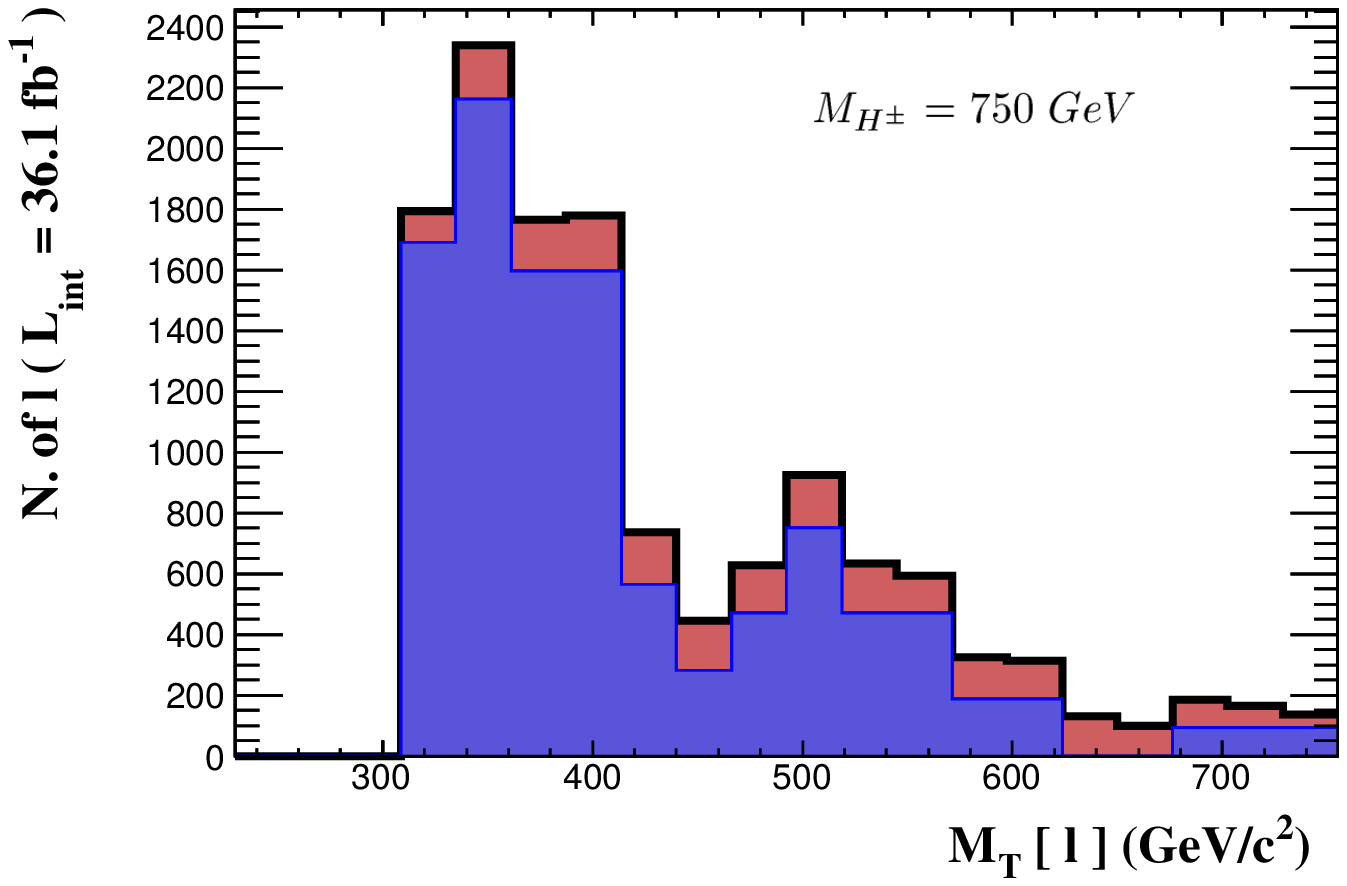}
\caption{Transverse mass plots for signal and background with all cuts taken into account, i.e., limited to the region used for the calculation of the significances.}
\label{pptaunu-cut-9}
\end{figure}

\begin{figure}
\includegraphics[scale=0.3]{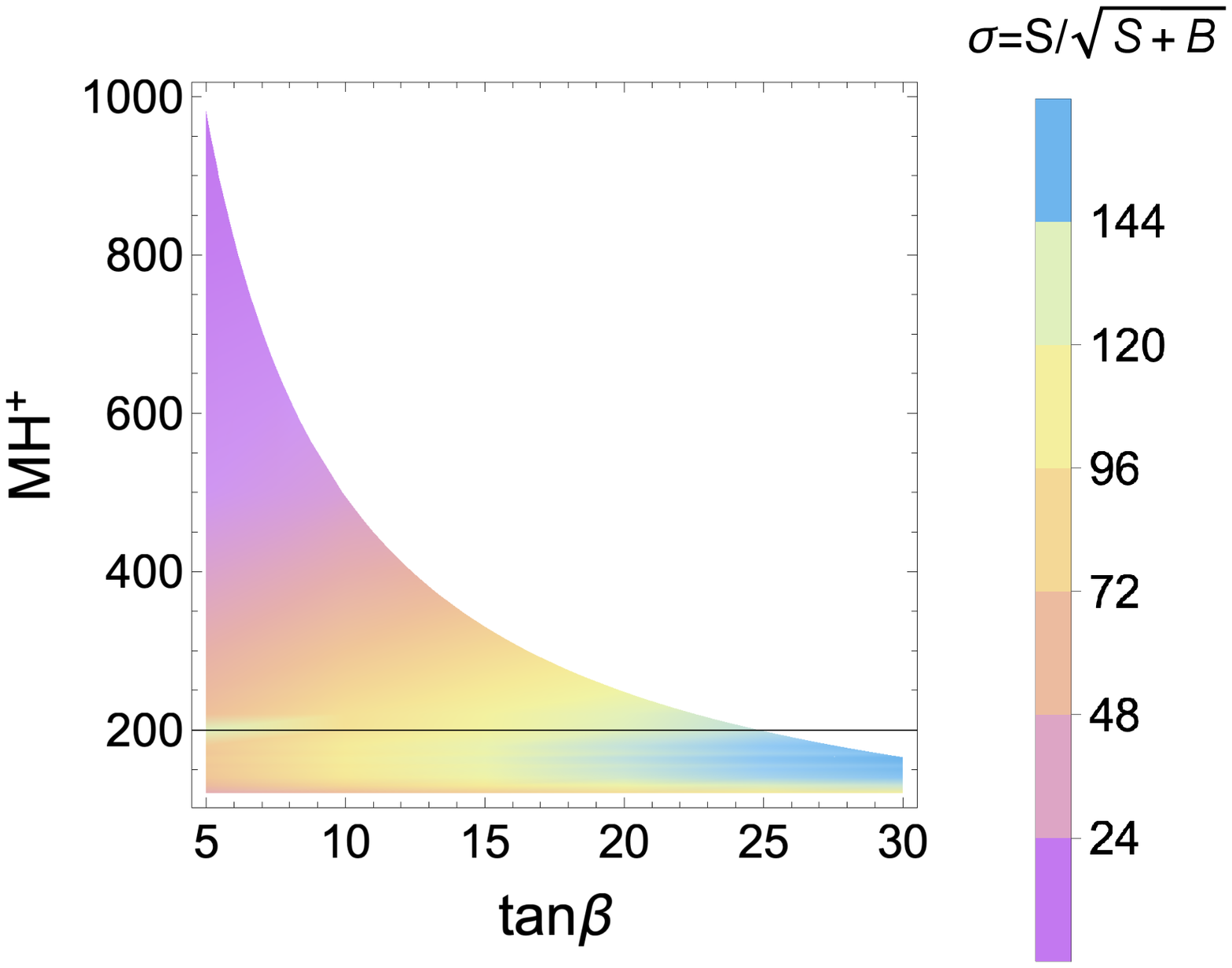}
\includegraphics[scale=0.3]{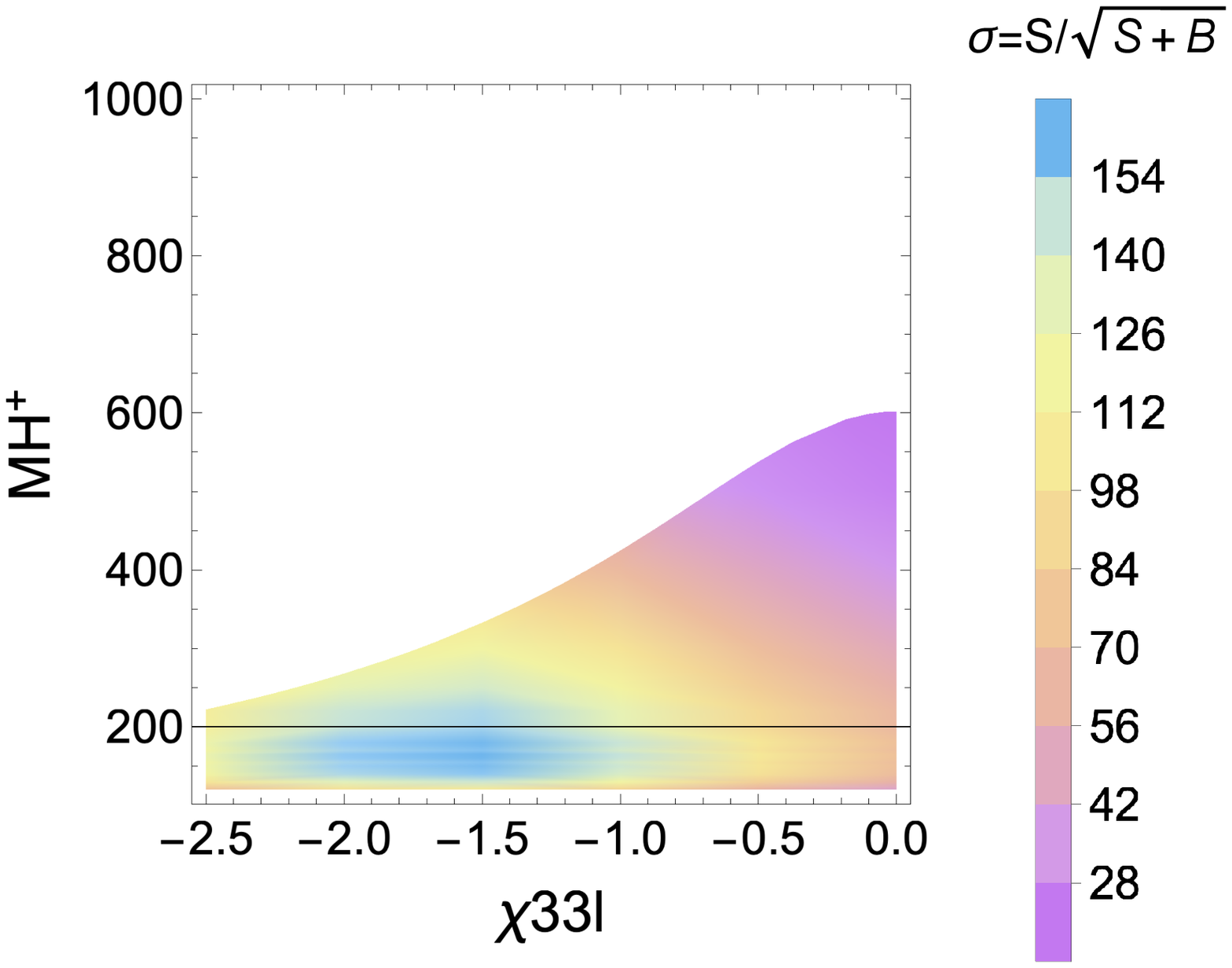}
\includegraphics[scale=0.3]{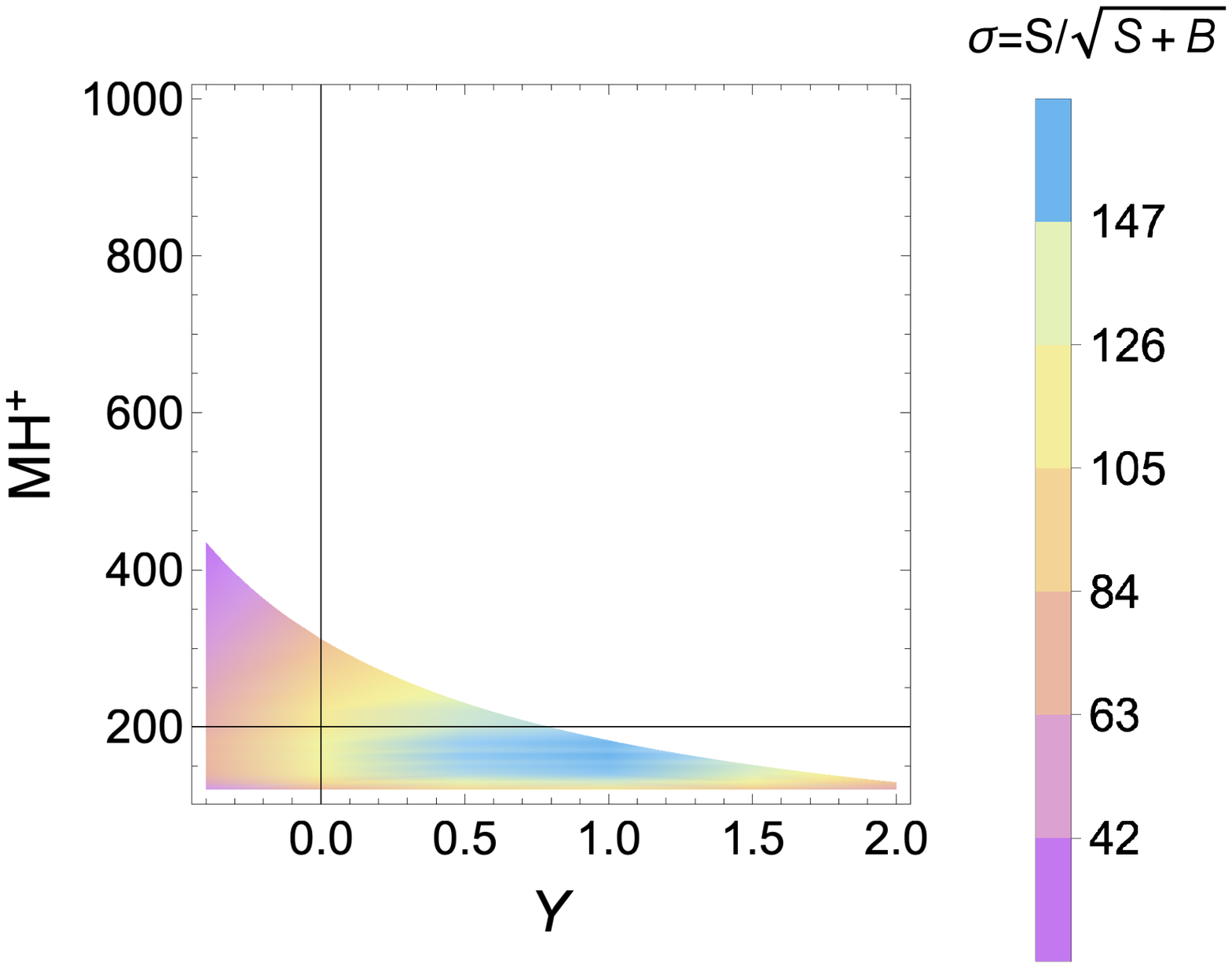}

\caption{Significance of our signal in terms of the most relevant parameters of the 2HDM-III like-X scenario.}
\label{pptaunu-cut-10}
\end{figure}

Following the above sequence of cuts, for which the signal and background responses can be found in Tab.~II, we revisit in Fig.~\ref{pptaunu-cut-9} the transverse mass distributions in the relevant peak regions. From these, the significances given in Tab.~III can be extracted. In turn, from the latter, it can be concluded that the signal is strong enough to be detectable at the LHC over a very large mass range, covering both the light and heavy mass regime of the charged Higgs boson stemming from the 2HDM-III. In fact, by interpolating between the various charged Higgs boson masses used in the MC analysis, we can perform a continuous scan of the relevant 2HDM-III like-X parameter space surviving current theoretical and experimental limits and map the signal significances, obtained at $L=36.1$ fb$^{-1}$ via the above search channel,  in terms of the 2HDM-III  input parameters to which the latter is sensitive, i.e., $\tan\beta$, $\chi_{33}^l$ (via $Y$) and $M_{H^{\pm}}$. This is done in Fig.~\ref{pptaunu-cut-10}.

\begin{table} \label{pptaunu_cuts}
\begin{center}
\begin{tabular}{|c|c|c|c|c|c|}
\hline 
$M_{H^{\pm}}$ & Cut 1: $p_T(l)$ & Cut 2: $\slashed{E}_T$ & Cut 3:$| \eta(l)|$ & Cut 4: $E_{T}$ & Cut 5: $M_{T}(l)$ \\ 
\hline 
$120$ GeV & $ \geq 45$ GeV & $\geq 40$ GeV &  $ \leq 1.2$ GeV & $ \geq 55$ GeV & $\geq 85$ GeV \\ 
 &  & $\leq 70$ GeV  &  &   & $\leq 125$ GeV \\
\hline 
Signal events & $294136 $ & $237167$ & $215684$ & $85480$ & $82147$ \\ 
\hline 
Background events & $27527568$ & $7919086$ & $3832807$ & $1090294$ & $795470$ \\ 
\hline 
$170$ GeV & $ \geq 45$ GeV & $\geq 60$ GeV &  $ \leq 1.2$ GeV & $ \geq 55$ GeV & $\geq 90$ GeV \\ 
 &  & $\leq 90$ GeV &  &  & $\leq 175$ GeV \\
\hline 
Signal events & $290051$ & $138676$ & $124849$ & $114334$ & $113758$ \\ 
\hline 
Background events & $ 27527568$ & $ 1282301$ & $ 669345$ & $ 568972$ & $ 536547$ \\ 
\hline 
$200$ GeV & $ \geq 45$ GeV & $\geq 70$ GeV &  $ \leq 1.2$ GeV & $ \geq 60$ GeV & $\geq 110$ GeV \\ 
 &  & $\leq 105$ GeV  &  &  & $\leq 205$ GeV \\
\hline 
Signal events & $233230$ & $94175$ & $84981$ & $80777$ & $80453$ \\ 
\hline 
Background events & $ 27527568$ & $ 627241$ & $ 333826$ & $ 304128$ & $290406$ \\ 
\hline 
$400$ GeV & $ \geq 45$ GeV & $\geq 100$ GeV &  $ \leq 1.2$ GeV & $ \geq 80$ GeV & $\geq 170$ GeV \\ 
 &  & $\leq 225$ GeV &  &  &  $\leq 405$ GeV \\
\hline 
Signal events & $ 42612$ & $ 22833$ & $ 20864$ & $ 20714$ & $20578$ \\ 
\hline 
Background events & $ 27527568$ & $ 146801$ & $ 80449$ & $ 78475$ & $ 74904$ \\ 
\hline 
$500$ GeV & $ \geq 45$ GeV & $\geq 90$ GeV &  $ \leq 1.2$ GeV & $ \geq 75$ GeV & $\geq 200$ GeV \\ 
 &  & $\leq 270$ GeV &  &  & $\leq 505$ GeV \\
\hline 
Signal events & $20674$ & $14716$ & $13292$ & $13194$ & $12021$ \\ 
\hline 
Background events & $ 27527568$ & $ 238246$ & $ 129132$ & $ 124809$ & $ 71238$ \\ 
\hline 
$750$ GeV & $ \geq 45$ GeV & $\geq 105$ GeV & $ \geq 80$ GeV & $\geq 1$ GeV & $\geq 320$ GeV \\ 
 &  &  &  &  & $\leq 755$ GeV \\
\hline 
Signal events & $ 4381 $ & $ 3351$ & $ 3049$ & $ 3042 $ & $ 2279$ \\ 
\hline 
Background events & $ 27527568 $ & $ 124057 $ & $ 68043 $ & $ 66539$ & $ 10714$ \\ 
\hline 
\end{tabular} 
\caption{Number of events after doing the multiplicity cuts of signal and background each cut described in the text, adopting the same sequence,  {{$L=36.1$ fb$^{-1}$}}. }
\end{center}
\end{table}

\begin{table} \label{pptaunu_sigmas}
\begin{center}
\begin{tabular}{|c|c|c|c|}
\hline 
$H^{\pm}$ mass (GeV) & Signal & Background & $S/\sqrt{S+B}$ \\ 
\hline 
120 & 82147 & 795470 & 87.688 \\ 
\hline 
130 & 111026 & 745095 & 119.994 \\ 
\hline 
140 & 138553 & 852330 & 139.189 \\ 
\hline 
150 & 133205 & 719156 & 144.282 \\ 
\hline 
155 & 123148 & 633444 & 141.578 \\ 
\hline 
160 & 131734 & 673010 & 146.849 \\ 
\hline 
165 & 133767 & 683161 & 147.999 \\ 
\hline 
170 & 113758 & 536547 & 141.067 \\ 
\hline 
175 & 117716 & 544818 & 144.621 \\ 
\hline 
180 & 121355 & 566716 & 146.299 \\ 
\hline 
200 & 80453 & 290406 & 132.111 \\ 
\hline 
220 & 79475 & 292568 & 130.297 \\ 
\hline 
250 & 73119 & 314654 & 117.420 \\ 
\hline 
300 & 38855 & 112403 & 99.906 \\ 
\hline 
400 & 20578 & 74904 & 66.597 \\ 
\hline 
500 & 12021 & 71238 & 41.66 \\ 
\hline 
750 & 2279.4 & 10714 & 19.997 \\ 
\hline 
800 & 1643.8 & 9586.2 & 15.511 \\ 
\hline 
1000 & 637.5 & 5263 & 8.299 \\ 
\hline 
\end{tabular} 
\caption{Significances after the complete sequence of cuts described in the text with $L=36.1$ fb$^{-1}$. }
\end{center}
\end{table}

\section{Conclusions}

In summary, there exist significance chances to extract a charged Higgs boson signal at the LHC within the 2HDM-III scenario in its like-X incarnation, by searching for the production and decay channel  $b \bar{c} \to H^-  \to \tau \bar \nu_\tau$, wherein the $\tau$ is identified through its  transitions into electrons/muons and corresponding neutrinos (the latter yielding transverse missing energy). This can be achieved by the end of Run 3 over a $H^\pm$ mass  interval ranging from 100 GeV or so up to the TeV scale. In order to obtain this, a dedicated selection procedure is required to be optimised around a tentative charged Higgs boson mass value. We have proven this to be very effective against the (dominant) background given by  $b \bar{c} \to W^-  \to \tau \bar \nu_\tau$ 
as well as the (subdominant) noise produced via $g {q}'  \to W^{\pm} q$ and  $q \bar{q}'  \to W^{+} W^{-} \to l^{+}l^{-}\nu \nu$. Finally, we are confident that our results are realistic, as we have obtained these through a sophisticated MC analysis exploiting advanced computational tools. We are therefore looking forward to ATLAS and CMS adopting our recommended approach, so as to confirm or disprove the 2HDM-III hypothesis. 

\section*{Acknowledgements}
SM is financed in part through the NExT Institute and the UK STFC Consolidated grant ST/L000296/1.
 SM acknowledge support from 
 the H2020-MSCA-RISE-2014 grant no.  645722 (NonMinimalHiggs). SR-N thanks the University of Southampton as well as Carleton University for hospitality while parts of this work were completed. JH-S and CH have been supported by SNI-CONACYT
(M\'exico), VIEP-BUAP and  PRODEP-SEP (M\'exico)
under the grant `Red Tem\'atica: F\'{\i}sica del Higgs y del
Sabor'. SR-N acknowledges a scholarship from
CONACYT (M\'exico). We all thank Professor Heather Logan  for useful and constructive discussions in the beginning of this work.

\end{document}